\documentclass[usenatbib,aas_macros]{mn2e}
\usepackage{booktabs}
\usepackage{multirow}
\usepackage[dvipdfmx]{graphicx,color}
\usepackage{amsmath}
\usepackage{color}
\usepackage{float}

\title[Globular cluster evolution with a million stars]{The DRAGON simulations: globular cluster evolution with a million stars}  
\author[Long Wang et al.]{Long Wang$^{1,2}$\thanks{E-mail:long.wang@pku.edu.cn}, Rainer Spurzem$^{3,7,1}$, Sverre Aarseth$^{4}$, Mirek Giersz$^{5}$,
\newauthor{Abbas Askar$^{5}$, Peter Berczik$^{3,7,8}$, Thorsten Naab$^{6}$, Riko Schadow$^{6}$}
\newauthor{and M.B.N. Kouwenhoven$^{1,2}$}\\
  $^{1}$Kavli Institute for Astronomy and Astrophysics, Peking University, Yiheyuan Lu 5, Haidian Qu, 100871, Beijing, China\\
  $^{2}$Department of Astronomy, School of Physics, Peking University, Yiheyuan Lu 5, Haidian Qu, 100871, Beijing, China\\
  $^{3}$National Astronomical Observatories and Key Laboratory of Computational Astrophysics, Chinese Academy of Sciences, \\
  20A Datun Rd., Chaoyang District, 100012, Beijing, China\\ 
  $^{4}$Institute of Astronomy, University of Cambridge, Cambridge, UK \\
  $^{5}$Nicolaus Copernicus Astronomical Centre, Polish Academy of Sciences, ul. Bartycka 18, 00-716Warsaw, Poland \\
  $^{6}$Max-Planck Institut f\"ur Astrophysik, Karl-Schwarzschild-Str. 1, D-85741 Garching, Germany \\
  $^{7}$Astronomisches Rechen-Institut, Zentrum f\"ur Astronomie, University of Heidelberg, M\"onchhofstrasse 12-14, \\
69120, Heidelberg, Germany\\
  $^{8}$Main Astronomical Observatory, National Academy of Sciences of Ukraine, 27 Akademika Zabolotnoho St., 03680, \\
Kyiv, Ukraine\\
}

\begin{document}

\date{Accepted 2016 February 1. Received 2016 February 1; in original form 2015 November 2}

\pagerange{---} \pubyear{---}

\maketitle

\label{firstpage}

\begin{abstract}
Introducing the DRAGON simulation project, we present direct $N$-body simulations of four massive globular clusters (GCs) with $10^6$ stars and 5$\%$ primordial binaries at a high level of accuracy and realism.
The GC evolution is computed with NBODY6++GPU and follows the dynamical and stellar evolution of individual stars and binaries, kicks of neutron stars and black holes, and the effect of a tidal field.
We investigate the evolution of the luminous (stellar) and dark (faint stars and stellar remnants) GC components and create mock observations of the simulations (i.e. photometry, color-magnitude diagrams, surface brightness and velocity dispersion profiles).
By connecting internal processes to observable features we highlight the formation of a long-lived 'dark' nuclear subsystem made of black holes (BHs), which results in a two-component structure.
The inner core is dominated by the BH subsystem and experiences a core collapse phase within the first Gyr. 
It can be detected in the stellar (luminous) line-of-sight velocity dispersion profiles. 
The outer extended core - commonly observed in the (luminous) surface brightness profiles - shows no collapse features and is continuously expanding.
We demonstrate how a King (1966) model fit to observed clusters might help identify the presence of post core-collapse BH subsystems. 
For global observables like core and half-mass radii the direct simulations agree well with Monte-Carlo models.
Variations in the initial mass function can result in significantly different GC properties (e.g. density distributions) driven by varying amounts of early mass loss and the number of forming BHs. 

\end{abstract}


\section{Introduction}
\label{sec:int}
It is one of the grand challenges of theoretical astrophysics to understand the evolution and dynamics of globular clusters (GCs).
Direct $N$-body simulations are challenging because typical GCs observed in the Milky Way have half-mass radii ($R_{\rm h}$) about $2$-$10$~pc with a short relaxation time ($\sim$~Gyr), and their mass can be higher than $10^5~M_{\rm \odot}$. 
The short timescale of binary interactions in GCs also increases the difficulty.
Since the early 90's \citep{Sugimoto1990} a million-body direct simulation including all required features (stellar evolution, binary interaction and external potential from host galaxies) has been considered a watershed mark for a complete simulation of GCs.
The important stepping stones were simulations of $10$k \citep{Spurzem1996}, $33$k \citep{Makino1996}, $100$k \citep{Baumgardt2003}, $200$k \citep{Hurley2012}, $263$k \citep{Sippel2013} and $485$k \citep{Heggie2014} stars.
Employment of new hardware and software significantly improved the performance of the simulation programs.
Essential was the series of NBODY codes developed by Aarseth and collaborators \citep{Aarseth2003, Nitadori2012, Wang2015} combined with the special-purpose hardware, first GRAPE \citep{Makino2003} and later on GPUs \citep{Gaburov2009,Nitadori2012,Wang2015}.

Various physical processes, that have to be considered simultaneously, pose an additional challenge.
First, GCs are gravothermal, i.e., the cumulative effect of distant gravitational encounters is important.
It is not possible to make a cutoff at a certain impact parameter, so in principle all pairwise gravitational interactions in the system need to be taken into account \citep{Spitzer1987}. 
This leads to an asymptotic scaling of the computational effort per dynamical time of $N^{7/3}$ with the particle number $N$ \citep{Makino1988}.
Efficient codes can significantly reduce the computing time needed for a certain problem \citep{Makino1988}, but the asymptotic scaling cannot be easily overcome, though some hybrid codes \citep[e.g.][]{Iwasawa2015,Rodriguez2015} may help in the future.
In practice even higher power of $N$ is found, because the relaxation time in physical units increases with $N/\log(N)$ (e.g. \citealp{Spitzer1987}).
On the other hand, astrophysically interesting GC particle numbers ($10^6$ or $10^7$) are not large enough for the reliable use of methods of statistical physics. 
Such methods, generally based on the Fokker-Planck equation, have been used with some success (e.g. \citealp{Giersz1994,Kim2008}), but when trying to include more physical processes for the simulation of real clusters they become too complex or even impossible to handle. 


There are many astrophysical processes involved in GC simulations.
Stellar evolution of individual stars leads to mass loss and the amount depends on the initial mass of a star and its chemical composition.
Also, a significant fraction of GC stars is in hard binaries (hard means that their binding energy is much higher than the average kinetic energy of stars in a cluster).
These binaries can have binding energies comparable to the entire energy of the cluster.
Their dynamical interactions with single stars or other binaries can dominate cluster evolution during certain stages (e.g. \citealp{Heggie1975,Hills1975,Heggie2006,Fregeau2003,Fregeau2007}). 
The stellar evolution of such binaries can lead to Roche-lobe overflow, mass loss, common envelope phases and possible mergers (see the first implementations of all these processes in the $N$-body models of \citealp{Hurley2005}). 
The considerable number of blue stragglers in GCs can only partially be explained by stellar evolution of binaries - in clusters with high stellar densities blue stragglers might form by direct stellar collisions (e.g. \citealp{Davies2004,Hypki2013,Davies2015}). 
Besides, the compact objects (white dwarfs, neutron stars and black holes) are not only interesting for X-ray, radio and gravitational wave observations, but also have significant influence on the dynamical evolution of GCs (e.g. \citealp{Downing2012,Hurley2007,Breen2013,Morscher2015,Contenta2015,Giersz2015}).
It turns out that all these processes, including long-distance stellar relaxation (which triggers core collapse and mass segregation), binary dynamics, single and binary stellar evolution, stellar collisions, interact with each other and with the GC dynamics in a complex way.
They are sensitive to initial conditions in a self-regulating way.
This led some authors to discuss the ``ecology'' of star clusters \citep{PZ1997}. 
Tidal fields and the interaction with a possible central massive black hole are further dynamical complications (e.g. \citealp{Lamers2005,Hurley2007,PZ2004,Kruijssen2011,Gieles2011,Giersz2015}), not to speak of the recent detection of possible dynamically distinct populations (e.g. \citealp{Piotto2009,Gratton2012}).

In this context ``understanding'' GC evolution can only mean the generation of a large number of models with different initial conditions and trying to understand how to map the current conditions of clusters back to their initial state. 
Again, some studies with semi-analytical models (e.g. \citealp{Alexander2014,Pijloo2015}) and Monte-Carlo models have been presented (e.g. \citealp{Joshi2000,Heggie2008,Giersz2011,Leigh2013,Leigh2015,Chatterjee2013,Morscher2015,Giersz2015}).
One of the Monte-Carlo codes is the MOCCA (MOnte Carlo Cluster SimulAtor) code \citep{Giersz2013}, which can provide very fast dynamical evolution of GCs with detailed information of individual cluster members.
However, all Monte-Carlo codes require verification by direct high-accuracy simulations.

GC simulations are relevant beyond the Milky Way and the Local Group.
GCs populate disks and spheroids of all massive galaxies.
The central parts of galactic nuclei also have dense stellar systems surrounding the massive black holes which resemble GCs (see \citealp{Genzel2010} and references therein).
Tens of thousands of GCs populate the intergalactic space in galaxy clusters \citep{Peng2011}, and GCs may trace the earliest conditions at the time of galaxy formation. 
On the other hand, galactic mergers create populations of GCs, so the merging history of massive galaxies could be reflected in the properties of their GC systems \citep{Brodie2006}.

In this paper we present the first step towards realistic simulations of GCs with initially one million particles, each of which represents an individual star.
The simulations are part of the ``DRAGON GC simulation project''\footnote{http://silkroad.bao.ac.cn/dragon/} and include models of single and binary synthetic stellar evolution.
We consider all relevant physical processes with current knowledge of GCs in the simulations and explore whether realistic GCs can be produced after $12$~Gyr.
We also find a good reference GC with similar stellar density and total mass in one of our DRAGON models for a comparison with observational properties.

Last, but not least, as more realistic GC models (in the DRAGON project) become available in the future, combined with a large number of Monte-Carlo (MOCCA) models, we can learn about specific properties of individual (Galactic or Local Group) GCs from our simulation results.
Our DRAGON models can be used for inverse dynamical synthesis - by presenting current time observational data from our model clusters we can link observational features to the unseen characteristics of current GCs.
These observational data include Hertzsprung-Russell diagrams, the distribution of light and the motion of visible stars (surface brightness and velocity dispersion profiles).
Then the simulation results can provide information about the distribution and motion of dark objects in their centers (black holes, neutron stars, white dwarfs), the dynamics and fraction of unresolved binaries, and the necessary conditions for the initial state of the cluster (mass, concentration and rotation). 
A similar method has been used to model the Pal~14, Pal~4 \citep{Zonoozi2011,Zonoozi2014} and M4 \citep{Heggie2014} clusters.

This paper is organized as follows. In Section~\ref{sec:method}, we briefly describe the NBODY6++GPU and the MOCCA code for simulations and the COCOA code used to simulate observational data.
Then we provide the initial conditions for our models in Section~\ref{sec:init}.
Section~\ref{sec:res} shows the results of our simulations including the comparison to NGC~4372 observational data.
Finally, we draw our conclusion in Section~\ref{sec:con} and discuss our results in Section~\ref{sec:dis}.

\section{Methods}
\label{sec:method}

\subsection{Direct $N$-body code}

A well-known direct $N$-body simulation code designed for star clusters is NBODY6, developed by \cite{Aarseth2003}.
The most important feature is the accurate treatment of binary and close encounter dynamics, which are crucial physical processes in star clusters, by using the algorithms of \cite{Kustaanheimo1965} and chain regularization \citep{Mikkola1993}.
Later, a GPU-based parallelized version (NBODY6GPU), designed for desktop or single computer nodes, was developed by \cite{Nitadori2012}.
This new version makes star cluster simulations with $10^5$ stars possible on one node.
Then recently, \cite{Wang2015} presented an MPI parallelization based code NBODY6++GPU, which is an extension of NBODY6GPU and can be used across multiple nodes on supercomputers.
This technical development enabled the first million-body simulations of GCs.

These codes also involve the single and binary stellar evolution recipes described in \cite{Hurley2000} and \cite{Hurley2002}, and galactic tidal fields.
There are some differences in the treatment of velocity kicks for neutron stars (NSs) and black holes (BHs) when they form after supernova explosions (Section~\ref{sec:init}).

All simulations were performed on the ``Hydra'' GPU cluster of the Max-Planck Computing and Data Facility (MPCDF), Germany.
For each simulation, $8$ nodes with a total of $160$ Intel Xeon E5-2650 cores and $16$ NVIDIA K20m GPUs were used.

\subsection{Monte-Carlo code}

The MOCCA code \citep{Giersz2013} used for comparison to the star cluster simulations presented here is a Monte Carlo code based on H$\acute{e}$non's implementation \citep{Henon1971}, which was further improved by Stodolkiewicz in the early 80s \citep{Stodolkiewicz1986}.
This method can be regarded as a statistical way of solving the Fokker-Planck equation. 
The basic assumptions behind the Monte Carlo method are: 
(1) spherical symmetry, which makes it easy to quickly compute the gravitational potential and stellar orbits at any place in the system; and
(2) cluster evolution is driven by two-body relaxation, and the time step at each position in the system is proportional to the local relaxation time. 
These are the reasons why the Monte Carlo method is much faster than any direct $N$-body code. 

MOCCA is a heterogeneous code, composed of independent modules including single and binary star evolution \citep{Hurley2000,Hurley2002}, few-body scattering \citep{Fregeau2004} and escape processes in tidally limited clusters \citep{Fukushige2000}.
It is able to follow most of the important physical processes that occur during the dynamical evolution of star clusters.

The MOCCA code has been extensively tested against the results of $N$-body simulations (\citealp{Giersz2013,Heggie2014} and references therein). 
The agreement between these two different types of simulations is very good. This includes the global cluster evolution, mass segregation timescales, the treatment of primordial binaries (energy, mass and spatial distributions), and the numbers of retained NSs and BHs. 
The key assumption implemented in the MOCCA code that guarantees this agreement is that, throughout the entire cluster, the timescale for significant evolution is always a small fraction of the local relaxation time, but larger than the local crossing time.

\subsection{Simulating Observations of $N$-body models}
\label{sec:s2o}

For the purpose of this study, we utilize the COCOA (Cluster simulatiOn Comparison with ObservAtions) code \citep{Askar2014} to simulate observations of the DRAGON clusters. 
The COCOA code is being developed to extend numerical simulations of star clusters for direct comparisons with observations.
It uses snapshots produced by Monte-Carlo or $N$-body simulations.
The snapshots from the NBODY6++GPU code contains information about the positions, velocities and stellar parameters of all objects in the star cluster at a specific time during the cluster evolution.
COCOA projects numerical data from the snapshot of the star cluster onto the plane of the sky and provides a complete projected snapshot with magnitudes of all objects in the cluster. 
The COCOA code can create observational data in the form of FITS files using the projected snapshot. 
The distance to the cluster, exposure time, resolution and other instrumental specifications for the simulated observational data obtained from COCOA can be adjusted.

With this we can create synthetic observations of simulated star cluster models from ground- and space-based telescopes like the Hubble Space Telescope (HST). 
We can extract observable cluster parameters by calculating surface brightness profiles, velocity dispersion profiles and by fitting data to analytical models (e.g. \citealp{King1966} model). 
With COCOA it is possible to create observations and catalogues of all objects in the cluster with their respective magnitudes in different photometric filters, to construct color-magnitude diagrams for simulated star cluster models, and to study populations of specific stars and stellar remnants.

\section{Initial models}
\label{sec:init}
Direct $N$-body simulations of million-body GCs are still very time consuming even on GPU-based supercomputers.
Therefore the construction of realistic initial models is very important.
In the best case, the simulations directly produce a specific observed GC.
The physical timescale of direct $N$-body codes is sensitive to the half-mass radius (\citealp{Wang2015}).
Simulations are faster if a cluster has initially a larger half-mass radius $R_{\rm h}$.
Thus our DRAGON project starts with low-density models (first three models in Table~\ref{tab:init}).
We searched the \cite{Harris1996} GC catalog and found that NGC~4372 has similar mass and density compared to our models.
We choose it as a reference GC for setting some initial conditions and for comparison with our results.

\subsection{NGC~4372}
NGC~4372 has a large half-light radius $R_{\rm hl} = 3.91'$ ($6.60$~pc), an absolute $V$ band magnitude $M_{\rm V} = -3.76$~mag \citep{Harris1996} and dynamical mass $M_{\rm dyn} = 2\times 10^5 M_{\rm \odot}$ \citep{Kacharov2014}.
From \cite{Harris1996}, the observed projected core radius (luminous stellar core) $R_{\rm cl}$ is $1.75'$ ($2.95$~pc) and the distance of NGC~4372 to the Sun is $5.8$~kpc (the distance to Galactic center is $7.1$~kpc).
The distance modulus is $(m-M)_V = 15.0$~mag \citep{Harris1996} and the interstellar extinction $E(B-V) \approx 0.39$~mag \citep{Piotto2002}.
For the analysis shown below, the apparent magnitude and color of stars are corrected using these two parameters when necessary.
This cluster is on a non-circular orbit around the Galactic center with an eccentricity of $0.45$ and a pericenter distance of $2.8$~kpc \citep{Casetti2007}.
\cite{Kacharov2014} observed the radial velocities of $131$ stars in NGC~4372 and found the cluster is rotating with an amplitude of $1.2$~km/s. 
It is an old ($15 \pm 4$~Gyr; \citealp{Alcaino1991}) and very metal-poor GC with a metallicity of $Z \approx 0.00016$ \citep{Geisler1995,Kacharov2014,San2015}. 
We use this metallicity value in our simulations.

\subsection{Initial model for $N$-body simulations}

We carry out four general simulations with realistic initial properties based on the current knowledge of GCs.
The fast Monte-Carlo (MOCCA) simulations were performed with the same initial conditions for comparison.
The models are named ``DRAGON'' clusters (D1, D2, D3 and D4 respectively).
To identify the key initial properties of the models, we use the following naming convention,``D[N]-R[A]-[B]'' (where ``[N'' is the index of DRAGON clusters, ``[A]'' represents the initial half-mass radius and ``[B]'' represents typical features of the models), instead of ``DRAGON$1$-$4$'' in the description of results.
Table~\ref{tab:init} shows the different initial parameters for these models.

\begin{table}
  \caption{The differences of initial models. $R_{\rm h,0}$ is the initial half-mass radius; $q$ denotes the initial binary mass ratio distribution; $Kick$ is the kick velocity model of NSs; $R_{\rm t,0}$ is the initial tidal radius.}
  \label{tab:init}
  \begin{tabular}{lllll}
    \hline\hline
                 & D1-   & D2-   & D3- & D4- \\
                 & R7-IMF93 & R7-IMF01 & R7-ROT & R3-IMF01  \\\hline
    Profile      & KW6$^{1}$ & KW6       &  ES6$^{2}$  & KW6        \\
    $R_{\rm h,0}$~(pc)    & 7.5       & 7.6      & 8.1        & 3.0      \\
    IMF          & IMF93$^{3}$     & IMF01$^{4}$     & IMF01   & IMF01 \\
    $q$          & RP$^{5}$        & K$^{6}$         & K       & K \\
    Kick         & Low$^{7}$       & High$^{8}$      & High    & High \\\hline
    $R_{\rm t,0}$~(pc) & 89       & 97       & 97         & 97 \\\hline
  \end{tabular}
  {\tiny
  \begin{tabular}{ll}
      $^{1}$ KW6:& \cite{King1966} ($W_0 = 6$)\\
      $^{2}$ ES6:& \cite{Einsel1999} ($W_0 = 6$; $\omega_0 = 0.8$) \\
      $^{3}$ IMF93:& \cite{Kroupa1993}\\
      $^{4}$ IMF01:& \cite{Kroupa2001} \\
      $^{5}$ RP:& Random pairing from IMF \\
      $^{6}$ K:& \cite{Kouwenhoven2007} (a distribution of $0.6 q^{-0.4} (0<q<1)$) \\
      $^{7}$ Low: & $\sigma_{\rm k} \approx 30$~km/s.\\
      $^{8}$ High: & $\sigma_{\rm k} = 265$~km/s.\\
  \end{tabular}
  }
\end{table}

The density of the non-rotational models (D1-R7-IMF93, D2-R7-IMF01 and D4-R3-IMF01) follows a spherical \cite{King1966} model with a scaled central potential parameter $W_{\rm 0} = 6$.
The initial half-mass radii $R_{\rm h,0}$ of D1-R7-IMF93 and D2-R7-IMF01 are $7.5$-$7.6$~pc (low density models).
The \cite{King1966} model is a single-mass, isotropic dynamical model for star clusters based on steady-state solutions of the Fokker-Planck equation.
It includes the tidal cutoff of a galactic potential. 
The $W_{\rm 0}$ denotes the central concentration of the clusters.
The rotational model D3-R7-ROT follows a rotational King model \citep{Einsel1999} with $W_0 = 6$ and a rotation parameter $\omega_0 = 0.8$ and $R_{\rm h,0} = 8.1$~pc.
Because the $60\%$ Lagrangian radius of the rotational King model is constant when only $\omega_0$ is varied, we keep the same $60\%$ Lagrangian radius of D3-R7-ROT as D2-R7-IMF01.
Finally, the model D4-R3-IMF01 has a higher density with $R_{\rm h,0}=3.0$~pc.

All clusters are initialized with $N = 1.05\times10^6$ stars sampled from two initial mass functions (IMFs): the \cite{Kroupa1993} IMF (hereafter IMF93) in D1-R7-IMF93 and the \cite{Kroupa2001} IMF (hereafter IMF01) in D2-R7-IMF01, D3-R7-ROT and D4-R3-IMF01.
IMF93 is a three-component power-law distribution of the form $\xi (m) \propto m ^{-\alpha}$ and IMF01 has two components instead of three.
For the mass range $0.08 < m \le 0.5 M_{\rm \odot}$, we use $\alpha_{\rm 1} \approx 1.3$ for both IMFs.
The difference of these IMFs occurs at the high-mass end. 
IMF93 has $\alpha_{\rm 2} \approx 2.2$ in the mass range $0.5 \le m \le 1 M_{\rm \odot}$ and $\alpha_{\rm 3} \approx 2.7$ for mass $> 1 M_{\rm \odot}$, 
while IMF01 has $\alpha_{\rm 2} \approx 2.3$ for mass $> 0.5 M_{\rm \odot}$.
IMF01 is top-heavy compared to IMF93.
Thus it results in more massive stars in the initial conditions of our GCs.
In all models, the stellar mass range is from $0.08$ to $100~M_\odot$.

Due to the computing performance limit, we cannot use a primordial binary fraction close to unity, as suggested by \cite{Kroupa1995}.
Similar to previous studies \citep{Eggleton1989,Hurley2012,Heggie2014}, we assume a primordial binary fraction of $5\%$ ($50,000$ binaries) with a log-normal semi-major axis distribution from $0.005$ to $50$~AU and a thermal distribution of eccentricities.
We use the upper limit $50$~AU, which excludes soft binaries that would be quickly broken up if included.
Thus our binary fraction represents a larger binary fraction if all separations had been allowed (as compared to a field star distribution).
The mass ratio ($q$) distribution of IMF01 models follows \cite{Kouwenhoven2007} motivated by the observed trend for massive members to have high $q$\footnote{The mass of the first component is sampled from the IMF, and the mass of the second component is obtained from the proportionality $0.6 q^{-0.4}$ (for $0<q \le 1$).}.
For the IMF93 model we generate binaries by random pairing.

Theoretical studies indicate that NSs and BHs can receive high velocity kicks at formation due to anisotropic supernova explosions \citep{Fryer2004}.
\cite{Hobbs2005} observed the proper motion of field pulsars in the Milky Way and concluded that the initial kick velocities of NSs follow a Maxwellian distribution with a velocity dispersion of $\sigma_{\rm k} \approx 265$~km/s (\texttt{High} kick model).
However, observations have discovered several NS X-ray binaries in GCs like 47~Tuc, M5 and M4 \citep{Manchester2005}, which suggests some NSs should have low kick velocities and remain bound.
The standard NBODY6 code has an option to use a Maxwellian distribution with much lower $\sigma_{\rm k}$ ($\sim 30$~km/s in the D1-R7-IMF93 model\footnote{The value is two times the velocity scaling factor based on original kick model in NBODY6, see \citealp{Aarseth2012}.}), aimed to keep $10\%$ NSs in the GCs (\texttt{Low} kick model).
There is no consensus about the NS kick velocity distributions and therefore we test both the \texttt{High} kick model (D2, D3 and D4) as well as the \texttt{Low} kick model (D1).

The BH kick velocity is different from that of a NS.
\cite{Belczynski2002} suggest that the BHs ($10$-$20~M_\odot$) formed after the supernova explosion undergo a fall-back phase which prevents high kick velocities. 
Thus GCs would more easily retain BHs.
In D1-R7-IMF93, an incomplete \cite{Belczynski2002} correction\footnote{Due to a bug in the NBODY6++GPU detected after a few Gyr of the R7-IMF93 simulation, the Carbon-Oxygen core mass ranging from $5$-$7.6~M_{\rm \odot}$ did not have the kick velocity reduction as in the \cite{Belczynski2002} model, which resulted in a slightly lower retention rate of BHs in the cluster.} of BH kick velocity based on the Maxwellian distribution with $\sigma_{\rm k} = 30~$km/s is used.
In the IMF01 models (D2, D3 and D4), the complete model is used.

Since the tidal fields of GCs have a large uncertainty, here we use a simple assumption that the GCs have a circular orbit with a point-mass potential ($7.1$~kpc to the galactic center; the same as the current distance of NGC~4372).
For all models, the initial tidal radius $R_{\rm t,0}$ is $89$-$97$~pc (the slight difference is due to the different total masses in these models).
Thus all models represent initial tidally underfilling GCs.

\section{Results}
\label{sec:res}
After $\sim 8,600$ hours computing time on the Hydra cluster, the D1-R7-IMF93 model reached $12$~Gyr,
while D2-D7-IMF01 spent $\sim 4,700$ hours and D3-R7-ROT had $\sim 4,500$ hours.
D4-R3-IMF01 reached $1$~Gyr with $2,900$ hours.
The variation of the computing speed is due to the different crossing timescales.
Table~\ref{tab:mn} shows the remaining total mass $M$ and number of stars $N$ (here one binary is counted only once) of the DRAGON clusters at different ages.
At $12$~Gyr, all R7 models still retain more than $\sim 70\%$ of the stars and have total mass of $2.5$-$2.9 \times 10^5 M_\odot$.
This indicates that our models reach the typical mass and number of stars observed in GCs.
Note that here we do not claim that our models can represent all GCs since they do not reach the central densities and half-mass radii of many GCs (e.g. the classical core-collapse GCs).

Our models are initial tidally underfilling clusters (where the outer cutoff of the cluster density distribution is smaller than the cluster tidal radius), and the R7 models have large initial half-mass relaxation timescale ($T_{\rm rh,0} \approx 7-8$~Gyr).
Thus the mass loss driven by two-body relaxation is slow and a high fraction of remaining stars is expected.
However, there are large differences in $N$ between D1-R7-IMF93 and D2-R7-IMF01 at $12$~Gyr.
To clarify the origin of this, the ratio of tidal radius and half-mass radius, $R_{\rm t}/R_{\rm h}$, is shown in Table~\ref{tab:mn}.
For the \cite{King1966} model with $W_{\rm 0}=6$, $R_{\rm t}'/R_{\rm h} \approx 5.6$. 
Here $R_{\rm t}'$ is the tidal radius defined in the King model, which is different from $R_{\rm t}$ based on the galactic potential. 
In the initial stage, all models have a ratio $R_{\rm t}/R_{\rm h} > 11.8$. 
The ratio $R_{\rm t}/R_{\rm h}$ for D1-R7-IMF93 is always larger than $5.6$, so it keeps the tidally underfilling condition while D2-R7-IMF01 reaches the tidally filling stage after $4$~Gyr (due to faster expansion, which will be discussed below).
Thus the escape rate of stars increases significantly for D2-R7-IMF01 after the tidally filling, which results in a higher mass loss rate.

\begin{table*}
  \centering
  \caption{Number of stars $N$ and total mass $M$ of GCs at different ages.}
  \label{tab:mn}
  \begin{tabular}{ccccccccccccccccc}
    \hline\hline
       &\multicolumn{3}{c}{D1-R7-IMF93} & \multicolumn{3}{c}{D2-R7-IMF01} & \multicolumn{3}{c}{D3-R7-ROT} & \multicolumn{3}{c}{D4-R3-IMF01} \\\hline
       $T$ [Myr]  &   $M$ [$M_\odot$]  &  $N$   & $R_{\rm t}/R_{\rm h}$  &   $M$  [$M_\odot$]  &  $N$  & $R_{\rm t}/R_{\rm h}$ &   $M$  [$M_\odot$]  &  $N$  & $R_{\rm t}/R_{\rm h}$ &   $M$  [$M_\odot$]  &  $N$  & $R_{\rm t}/R_{\rm h}$ \\\hline
       0 & 474603 & 999997 & 11.80 & 591647 & 1000000 & 12.79 & 591647 & 1000000 & 11.96 & 591647 & 1000000 & 32.14\\
       50 & 435810 & 997822 & 10.46 & 480562 & 994007 & 9.26 & 480518 & 994031 & 8.50 & 480831 & 993985 & 22.86\\
       100 & 422848 & 997113 & 10.08 & 457850 & 992578 & 8.68 & 457753 & 992622 & 8.02 & 458097 & 992607 & 21.31\\
       500 & 391229 & 996244 & 9.08 & 412177 & 986522 & 7.57 & 409096 & 978347 & 7.04 & 412880 & 991451 & 16.86\\
       1000 & 376681 & 993536 & 8.54 & 392165 & 974614 & 7.09 & 387066 & 959812 & 6.79 & 394901 & 989118 & 14.31\\
       2000 & 357899 & 984748 & 7.84 & 365727 & 948383 & 6.33 & 364791 & 943509 & 6.12 &   &   &  \\
       4000 & 337478 & 964011 & 7.08 & 334481 & 898150 & 5.46 & 344775 & 926013 & 5.28 &   &   &  \\
       6000 & 323426 & 942882 & 6.67 & 310774 & 848380 & 4.91 & 330144 & 902402 & 4.80 &   &   &  \\
       8000 & 312686 & 922355 & 6.38 & 290368 & 798496 & 4.53 & 316543 & 872637 & 4.50 &   &   &  \\
       10000 & 303830 & 902487 & 6.21 & 271034 & 747900 & 4.22 & 302396 & 836923 & 4.26 &   &   &  \\
       12000 & 295835 & 883075 & 6.04 & 252874 & 697482 & 3.97 & 287917 & 796586 & 4.04 &   &   &  \\\hline\hline
  \end{tabular}
\end{table*}

\subsection{Photometry}
\label{sec:pht}

\begin{figure*}
  \begin{tabular}{cc}
  \includegraphics[width=0.5\textwidth]{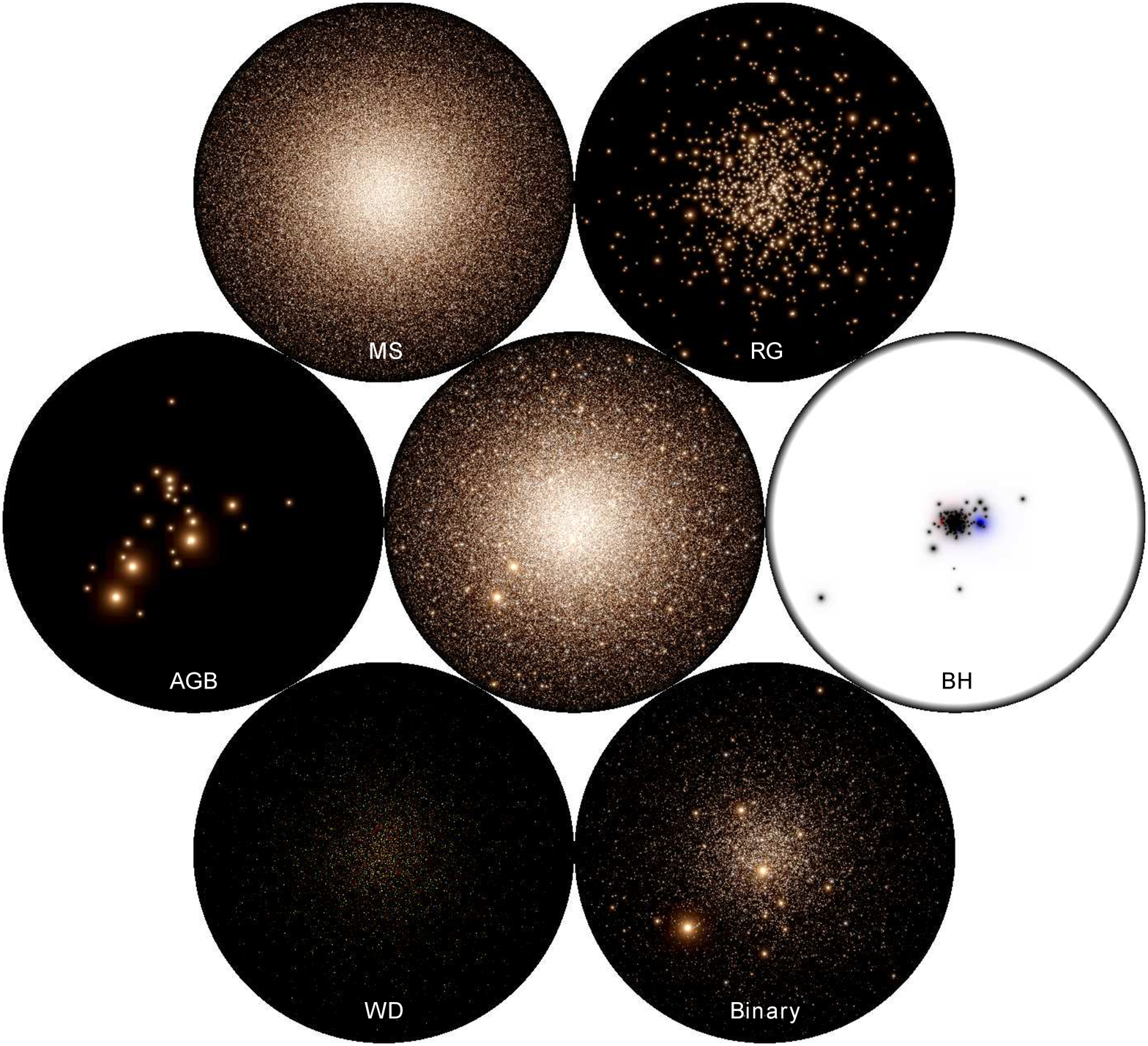} &
  \includegraphics[width=0.5\textwidth]{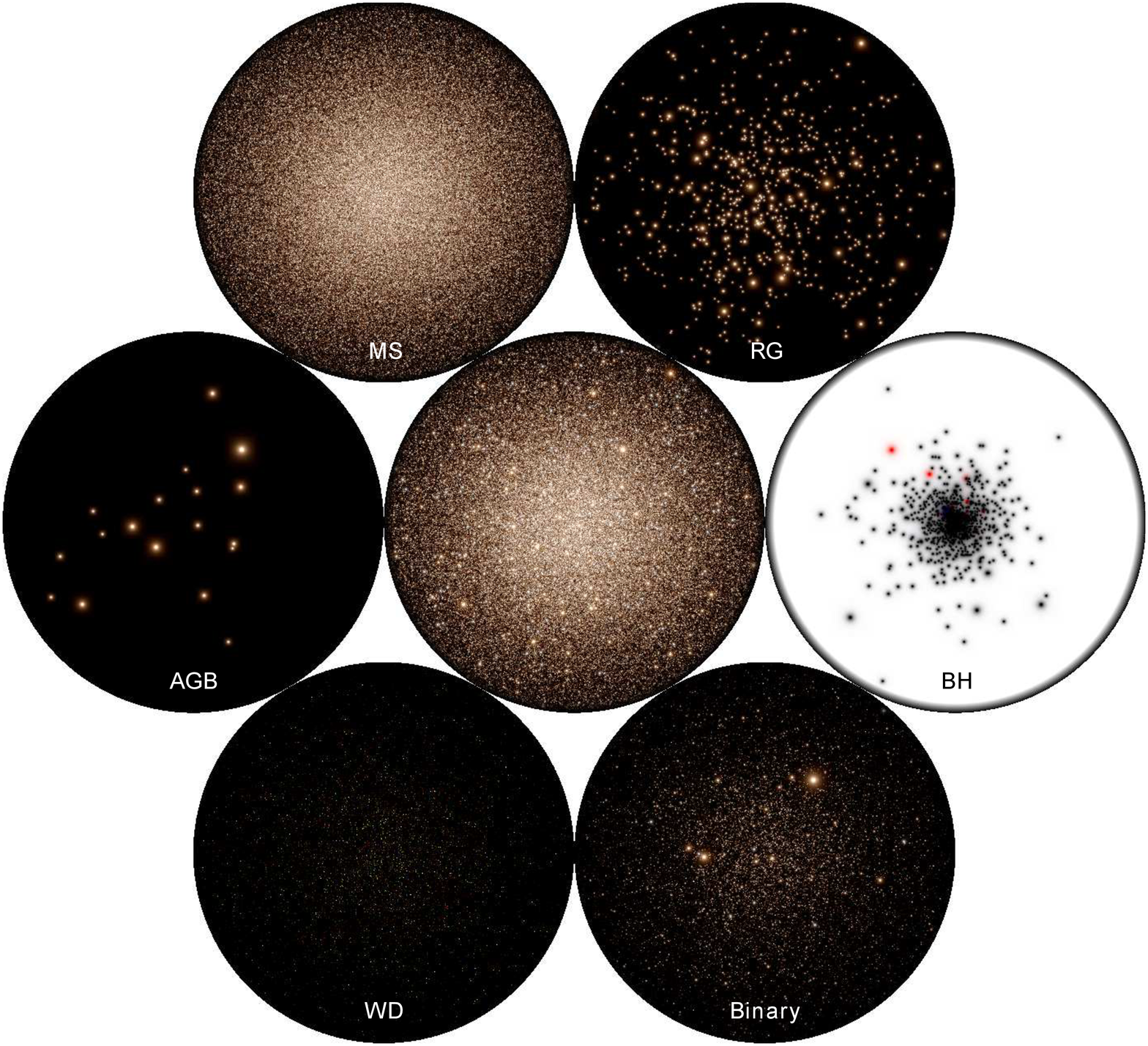}\\
  D1-R7-IMF93 & D2-R7-IMF01 \\
  \end{tabular}
  \caption{Snapshot of the D1-R7-IMF93 and D2-R7-IMF01 models at $12$~Gyr as observed in $B$, $V$ and $I$.
    The diameter of each image is $57.6$~pc with $2048\times2048$ pixels ($1.0$~arcsec/pixel). 
    The Moffat (1969) point spread function (PSF) is used here with seeing of $0.5$~arcsec.
    There is no background fluctuation, field stars contamination or cosmic rays.
    The colors are generated by $B$ (blue), $V$ (green) and $I$ (red) bands.
    The brightness is shown on a logarithm scale.
    The different types of stars are also shown individually. 
    The labels MS, RG, AGB, WD and BH represent main sequence, red giant, asymptotic giant branch, white dwarf and black hole respectively.
    The exposure time of white dwarfs is enlarged by a factor of $10^4$.
    The dot size in the BH panel is proportional to the BH mass; the red dots represent binaries with one BH component and the blue points are BH-BH binaries (due to crowding, particularly in the central regions, some binaries are overlapped by single BHs).
  }
  \label{fig:snap}
\end{figure*}

In general, simulations cannot be directly compared with observations due to the different parameter definitions and the observational sample incompleteness. 
To avoid this issue, we use the method described in Section~\ref{sec:s2o} to generate mock observations from our models.
This provides a way to ``observe'' DRAGON clusters by customized ``telescopes''.
First, by combining Johnson $B$ (blue), $V$ (green) bands and Cousins $I$ (red) bands ``photometry'' of simulation data, the images of D1-R7-IMF93 and D2-R7-IMF01 at $12$~Gyr are generated (Fig.~\ref{fig:snap}).
Special objects like red giants (RGs), asymptotic giant branch (AGB) stars, white dwarfs (WDs), BHs and binaries are also shown separately.
Although the contamination (e.g. cosmic rays, field stars and background fluctuations) is not included, Fig.~\ref{fig:snap} resembles real observed images of GCs.

A significant feature apparent in these images is the difference in concentration between the two models.
D1-R7-IMF93 is much denser centrally compared to D2-R7-IMF01.
The main difference between the initial conditions of these two models is the IMF (Table~\ref{tab:init}), which has a very strong influence on the dynamical evolution of GCs.
IMF01 models initially contain more massive stars than the IMF93 model.
This also results in very different numbers of retained BHs, as shown in Fig.~\ref{fig:snap}.
These BHs stay in the cluster center and form BH subsystems.
D1-R7-IMF93 has a centrally concentrated BH subsystem, while D2-R7-IMF01 has many more BHs by a factor $4$ and an extended structure.
Due to the small-number statistics, AGB stars show an asymmetric distribution.
In D2-R7-IMF01, even RGs, WDs and binaries have slightly asymmetric features.

\begin{figure}
  \centering
  \includegraphics[width=0.5\textwidth,height=!]{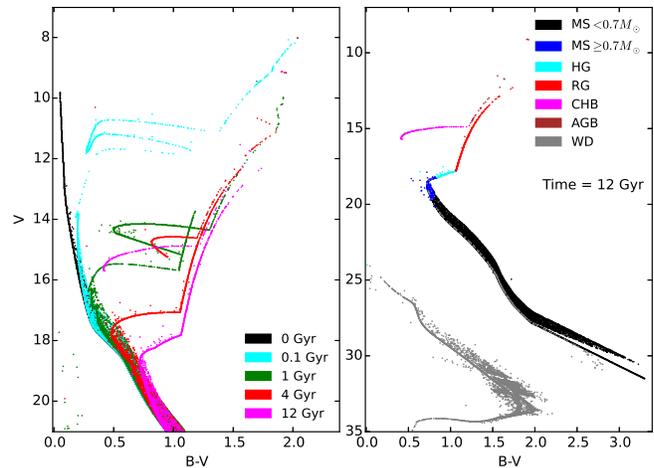}
  \caption{Color (B-V) - magnitude (apparent V) diagram (CMD) of D1-R7-IMF93. The distance modulus $13.82$ ($5800$~pc) is used here.
    Left: the evolution of CMD for luminous parts. Black dots show the initial distribution. After $100$ Myr the MS turnoff becomes visible already (cyan). The horizontal branch is populated after $1$~Gyr (green). The MS turnoff moves down to $V \approx 17.5$~mag after $4$~Gyr (red) and $V \approx 18.5$~mag after $12$~Gyr (purple).
    Right: the full CMD after $12$~Gyr. Different colors show different stellar types. HG: sub-giant branch (Hertzsprung gap); RG: red giant; CHB: horizontal branch (core helium burning); AGB: asymptotic giant branch; WD: white dwarf.}
  \label{fig:cmd}
\end{figure}

The direct view of these images shows no significant concentration difference between different stellar types within each model, except for the BHs.
The WDs are very faint but have a large color range (see also Fig.~\ref{fig:cmd}).

With the $B$, $V$ and $I$ band data, the color-magnitude diagrams (CMD) can be obtained immediately.
Fig.~\ref{fig:cmd} shows the CMD of D1-R7-IMF93 with the apparent $V$ magnitude (assuming a distance of $5.8$~kpc, similar to NGC~4372) and color $B-V$.
All binaries are treated as unresolved objects, and the binary broadening can be seen in the MS branch.
The left panel of Fig.~\ref{fig:cmd} shows the evolution of the CMD.
Initially, all stars are on the MS branch.
After $100$~Myr, the MS turnoff is visible and very bright stars with $V<10$~mag appear.
Then the MS turnoff significantly decreases to $V \approx 15$~mag at $1$~Gyr.
After $4$~Gyr, it moves to $V \approx 17.5$~mag with redder color and reaches $V \approx 18.5$~mag after $12$~Gyr.
The horizontal branch (core helium burning; CHB) is already populated at $1$~Gyr and becomes narrow in color after $4$~Gyr.
Then it broadens again after $12$~Gyr.

The right panel in Fig.~\ref{fig:cmd} shows the distribution of different stellar components after $12$~Gyr.
There are a few AGB stars with three very bright ones ($V<10$~mag). 
The MS turnoff mass is close to $0.8~M_\odot$.
Above the sub-giant branch (Hertzsprung gap; hereafter HG), the stars are distributed in a regular way without broadening.

The WDs follow the well-known ``cooling track'' behaviour with a hook feature that results from the cooling rate decreasing with age.
The locus of the track depend on the mass of WDs.
In Fig.~\ref{fig:cmd}, the WDs are scattered around the hook ($V\approx 30$~mag, $B\approx 1.5$~mag).
A part of the scatter is induced by the WDs formed from interacting binaries.
But due an inconsistent treatment of some WD ages\footnote{In D1-R7-IMF93 only, the duration of a star staying on the AGB stage before evolving to WD is too long (about $100$~Myr) for the first few hundred Myr simulation, thus some WDs are assigned wrong ages. The luminosity and stellar radius of these WDs were calculated incorrectly in the simulation. We re-calculate them by assuming that these WDs have an age of $12$~Gyr.}, some of the scatter is artificial.
Fortunately, the WDs are very faint and will not affect our main results.

\begin{figure}
  \centering
  \includegraphics[width=0.5\textwidth,height=!]{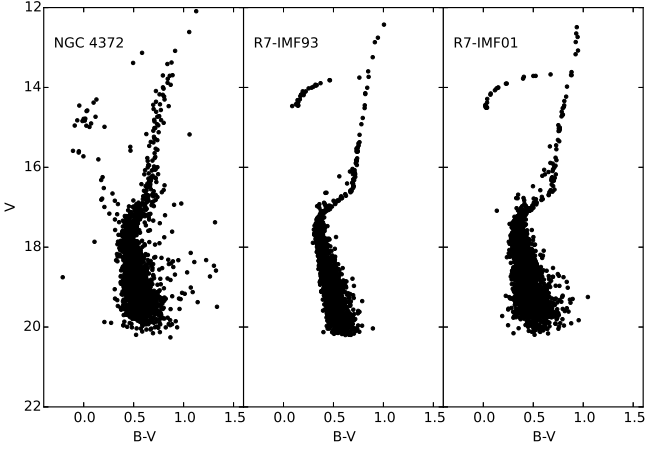}
  \caption{The CMD of NGC~4372 from HST observations (Piotto et al., 2002) and the simulated observation-like CMD of D1-R7-IMF93 and D2-R7-IMF01 at $12$~Gyr. }
  \label{fig:cmdobs}
\end{figure}

Fig.~\ref{fig:cmd} demonstrates that our simulations can recover all major features that are typical for observed single-population GCs.
But it is the ideal ``observation'' of the CMD with no selection bias and photometry errors. 
For a better comparison with real observational data, we generate the HST-like photometry of D1-R7-IMF93 and D2-R7-IMF01 at $12$~Gyr using the COCOA tool, followed by the standard observational data reduction.
The mock CMDs (middle and right panels of Fig.~\ref{fig:cmdobs}) agree well with HST observations of NGC~4372 \citep{Piotto2002} (left panel of Fig.~\ref{fig:cmdobs}).
For NGC~4372, the $B$ magnitude is converted from the F439W filter and $V$ is from the F555W filter by using the formalism described in \cite{Piotto2002}.
The photometry is obtained from the HST WFPC2 camera and the extinction is corrected.

In our models, we use a similar pixel scale ($0.0455$~arcsec/pixel) and pixel size ($1600\times1600$) as WFPC2 (excluding the PC chip).
The \cite{Moffat1969} point spread function (PSF) is used with seeing of $0.1$~arcsec.
Since our models have different half-light $R_{\rm hl}$ and core radii $R_{\rm cl}$ from NGC~4372 (D1-R7-IMF93 has $R_{\rm hl} \approx 8.7~$pc, D2-R7-IMF01 has $R_{\rm hl} \approx 14.4$~pc and NGC~4372 has $R_{\rm hl} \approx 6.6~$pc), it is difficult to compare the images at the same distance to the Sun ($R_{\rm G}=5800$~pc).
In this case, the observed number of stars and the projection effects would be very different (projection effect depends on the distance to the cluster center).
The mass segregation will also influence the results.
Thus, we keep the pixel scaling with two-dimensional luminous stellar core radii $R_{cl}$ of our models and NGC~4372 (discussed in Section~\ref{sec:radius}) by varying the $R_{\rm G}$, which means pixel/$R_{cl}$ ratio is the same for all three GCs.
The exposure time and detection limits are also adjusted to keep the total number of detected stars similar.
Although the CMDs from the simulations are very similar to the observed CMD, some differences are noticeable.
In our models, the RG and AG branches are narrow as compared to NGC~4372.
This is either caused by the observational photometry error or by the presence of multiple populations in NGC~4372 (our simulations have one single population and large spread is also detected in \citealp{San2015}).
Another difference is that our models have a continuous CHB which is not seen in NGC~4372.
This is caused by the stellar evolution algorithm in combination with the smooth IMF used in our simulations.
Another possibility is that in NGC~4372, some CHB stars might be ejected by binary/dynamical interactions and this process is not efficient enough in our models due to lower density or the different binary properties.

\begin{figure}
  \centering
  \includegraphics[width=0.5\textwidth,height=!]{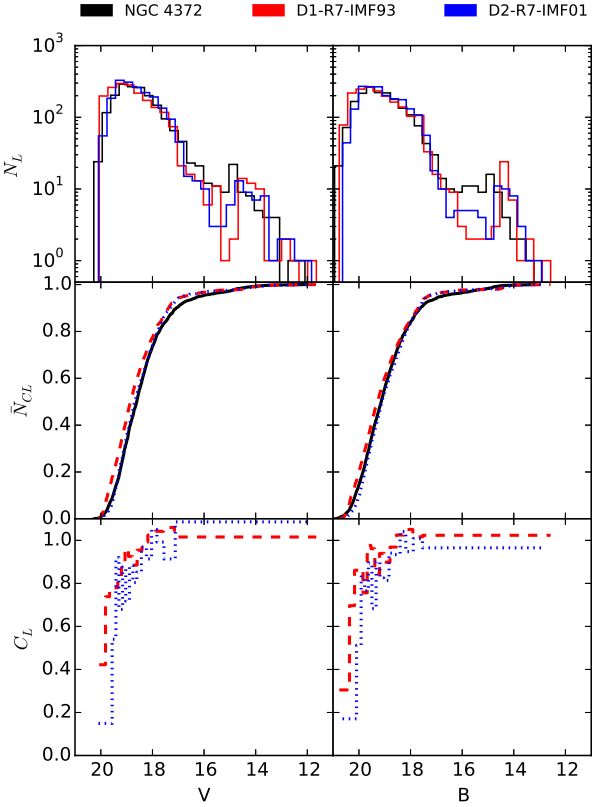}
  \caption{The luminosity distribution of stars in NGC~4372 from HST observation (Piotto et al., 2002) and D1-R7-IMF93 (red) and D2-R7-IMF01 (blue) at $12$~Gyr. The upper panels ($N_{\rm L}$) are the absolute luminosity distributions of $V$ and $B$ bands. The middle panels ($\bar{N}_{\rm CL}$) show the normalized cumulative luminosity distributions and the lower panels ($C_{\rm L}$) show the completeness per luminosity bin of our models.}
  \label{fig:histvb}
\end{figure}

In addition, our simulated photometric data not only provides realistic CMDs, but also have a similar luminosity distribution of stars $N_{\rm L}(M)$ as compared to NGC~4372.
Fig.~\ref{fig:histvb} shows the $N_{\rm L}(M)$ of D1-R7-IMF93, D2-R7-IMF01 and NGC~4372.
Except for the range of $V = 17$-$14$~mag, which corresponds to the CHB and a part of AGB shown in Fig.~\ref{fig:cmdobs}, $N_{\rm L}(M)$ of our models have an almost identical shape as NGC~4372.

The normalized cumulative luminosity distribution $\bar{N}_{\rm CL}(M)$ of all three GCs are also very similar.
The stars with $V<18$~mag have MS mass $>0.7 M_\odot$ with a turnoff mass of $\sim 0.8 M_\odot$ and AGB stars have masses $\sim 0.7 M_\odot$.
Thus $\bar{N}_{\rm CL}(M)$ suggests that there are no significant differences in the mass function up to $~0.8~M_\odot$ after $12$~Gyr for the IMF93 and IMF01 models.
Therefore, observations of the luminosity function of bright stars only is not sufficient to discriminate between IMFs, at least between those of IMF93 and IMF2001.

The completeness of the detected stars as a function of magnitude $C_L(M)$ can now easily be obtained (lower panels in Fig.~\ref{fig:histvb}).
$C_{\rm L}(M)$ grows to unity in the range from $20$ to $17$~mag in both the $B$ and $V$ band.
D1-R7-IMF93 and D2-R7-IMF01 have similar $C_{\rm L}(M)$ in this range.
Above $17$~mag, the $C_{\rm L}(M)$ of D1-R7-IMF93 is close to unity.
However, D2-R7-IMF01 has $C_{\rm L}(M)>1$ in the $V$ band and $C_{\rm L}(M)<1$ in the $B$ band.
The $N_{\rm L}(M)$ suggests that the number of stars above $17$~mag is very small, thus a contamination from other magnitude bins due to the photometric error can cause this scattering effect.

\subsection{Compact stellar remnants}

When massive stars evolve to compact objects (WDs, NSs or BHs), BHs become the most massive objects in the cluster, followed by the NSs and WDs.
They can dramatically change the dynamical evolution of star clusters.
The remaining number and cumulative number of newly-formed WDs, NSs and BHs at different ages are shown in Table~\ref{tab:wnb}.
Due to the fall-back models for BH initial kick velocity after supernova explosion (see Section~\ref{sec:init}), significant fractions of BHs remain bound to the clusters.
All R7 models reach $1.5$-$1.7~T_{\rm rh,0}$ after $12$~Gyr, so as expected from the recent studies of BH subsystems in star clusters (e.g. \citealp{Breen2013,Morscher2015}), the BH subsystem survives until the end of the simulation ($12$~Gyr).
This may not be the case for a model with much higher initial density with $T_{\rm rh,0} \approx 1~$Gyr.
Due to different IMFs, a total of $629$ BHs formed in D1-R7-IMF93 and about $2000$ formed in the IMF01 models.
At $12$~Gyr, D1-R7-IMF93 retains $245$ BHs and D2-R7-IMF01 and D3-R7-ROT have $\sim 1000$.
Here we notice that due to an inaccurate treatment of the binary stellar evolution look-up time, there are a few abnormal BHs with mass larger than $30 M_\odot$ formed within the first few Myr in D2-R7-IMF01 and D3-R7-ROT\footnote{The look-up time is the time point for the next stellar evolution check of one star. The problem was detected after $6$~Gyr of the simulations time of D2-R7-IMF01 and D3-R7-ROT. It causes incorrect look-up times of some binaries. Their stellar evolution is too fast. Some of these binaries finally merged and formed new stars with incorrect ages.  They became the progenitors of the abnormal BHs.}.
This is a similar issue to that detected and discussed for the model of \cite{Heggie2014}, which had further complications owing to the shorter relaxation time of the model in that work.
In our models, these BHs represent only a very small fraction of the total number of BHs and their influence is not significant.

The \texttt{Low} kick model of D1-R7-IMF93 results in about one hundred NSs retained at $12$~Gyr while other models have zero or few (D4-R3-IMF01 has one at $1$~Gyr).
One may notice that at $50$~Myr, D2-R7-IMF01 and D3-R7-ROT contain about $100$ NSs, while the denser model D4-R3-IMF01 only has $68$. 
This difference is caused by stochastic effects.
For a number of $6600$ NSs, we re-checked the possible number of NSs that can have initial kick velocities below the cluster escape velocity by random sampling from the Maxwellian velocity distribution with $\sigma=265$~km/s. 
The results show this number can vary from $0$ to $10$. 
We use the same random seed of the kick velocity generator for the first two models, thus they by chance have similar number of relatively low kick velocity NSs while D4-R3-IMF01 has less.
However, the velocities of these NSs are still larger than the cluster escape velocity. 
Thus all of them, except for two in D4-R3-IMF01, finally escaped.

In the MOCCA simulation with the same initial conditions as D2-R7-IMF01, about $10$ NSs are retained in the cluster even at $12$~Gyr.
This difference is also due to stochastic effects.
The MOCCA simulation might by chance have more NSs with low kick velocities.
In addition, a NS can also form from the accretion of WDs in a mass-transfering binary system. 
This channel can also result in a different number.
In our models, due to a technical issue in the current version of our code, the tidal circularization process of binaries is switched off.
This results in a smaller probability of NSs being formed through this channel (there was no detection of this event in our models).

The models with low kick velocities (D1-R7-IMF93) allow the NSs to remain bound to the GC during its evolution.
This is consistent with the fact that NS X-ray binaries have been directly observed \citep{Manchester2005}.
In the models with high kick velocities (D2, D3 and D4), which are consistent with the observed NS proper motions in \cite{Hobbs2005}, all NSs have escaped. 
This is in contradiction to the direct detections of NSs in GCs.
However, since our D2-R7-IMF01/ROT models have relatively large initial $R_{\rm h}$ (corresponding to escape velocity $\sim 30$~km/s), the NSs have more difficulty to remain bound in the GCs since most of them have kick velocities larger than the cluster's escape velocity. 

The total number of WDs formed within $12$~Gyr are similar for all R7 models. 
For the same IMF and initial number of stars, D2-R7-IMF01 has about $2000$ fewer WDs formed than D3-R7-ROT's.
The reason is due to more escapers from D2-R7-IMF01: the escapers which are not WDs during escape but could form WDs later are not counted.
There is no initial kick of WDs when they form, thus most WDs remain bound to the clusters.

\begin{table*}
  \centering
  \caption{Number of WDs, NSs and BHs remaining in the clusters. ``ALL'' means the total formed number including escapers. For D4-R3-IMF01, only the members up to $1$~Gyr are counted. Note that the escapers that were not WD, NS or BH, but could form later, were not counted.}
  \label{tab:wnb}
  \begin{tabular}{c|ccc|ccc|ccc|ccc}
    \hline\hline
    &\multicolumn{3}{c}{D1-R7-IMF93} & \multicolumn{3}{c}{D2-R7-IMF01} & \multicolumn{3}{c}{D3-R7-ROT} & \multicolumn{3}{c}{D4-R3-IMF01} \\\hline
    Time [Myr] & WD & NS & BH & WD & NS & BH & WD & NS & BH & WD & NS & BH\\\hline
    50 & 0 & 420 & 438 & 0 & 104 & 1436 & 0 & 106 & 1432 & 0 & 68 & 1458\\
    100 & 53 & 172 & 435 & 3232 & 0 & 1435 & 3228 & 1 & 1428 & 3204 & 3 & 1458\\
    500 & 15254 & 150 & 433 & 20783 & 0 & 1421 & 20642 & 0 & 1420 & 20821 & 2 & 1401\\
    1000 & 25681 & 143 & 425 & 32302 & 0 & 1409 & 31914 & 0 & 1414 & 32705 & 2 & 1329\\
    2000 & 43080 & 137 & 401 & 49311 & 0 & 1352 & 49237 & 0 & 1374 &  &  & \\
    4000 & 62740 & 130 & 348 & 66306 & 0 & 1258 & 68570 & 0 & 1282 &  &  & \\
    6000 & 75916 & 125 & 305 & 75836 & 0 & 1180 & 80522 & 0 & 1220 &  &  & \\
    8000 & 85323 & 125 & 281 & 81215 & 0 & 1130 & 88659 & 0 & 1172 &  &  & \\
    10000 & 92572 & 124 & 262 & 84354 & 0 & 1080 & 94250 & 0 & 1130 &  &  & \\
    12000 & 98563 & 121 & 245 & 85614 & 0 & 1037 & 97824 & 0 & 1096 &  &  & \\\hline
    ALL  & 104175 & 2942 & 629 & 105652 &6697& 2096& 107587 & 6696& 2096& 32732 &6736& 2092 \\\hline
  \end{tabular}
\end{table*}

\subsection{Spatial distribution of stellar components}
\label{sec:sc}

\begin{figure*}
  \centering
  \includegraphics[width=1.0\textwidth,height=!]{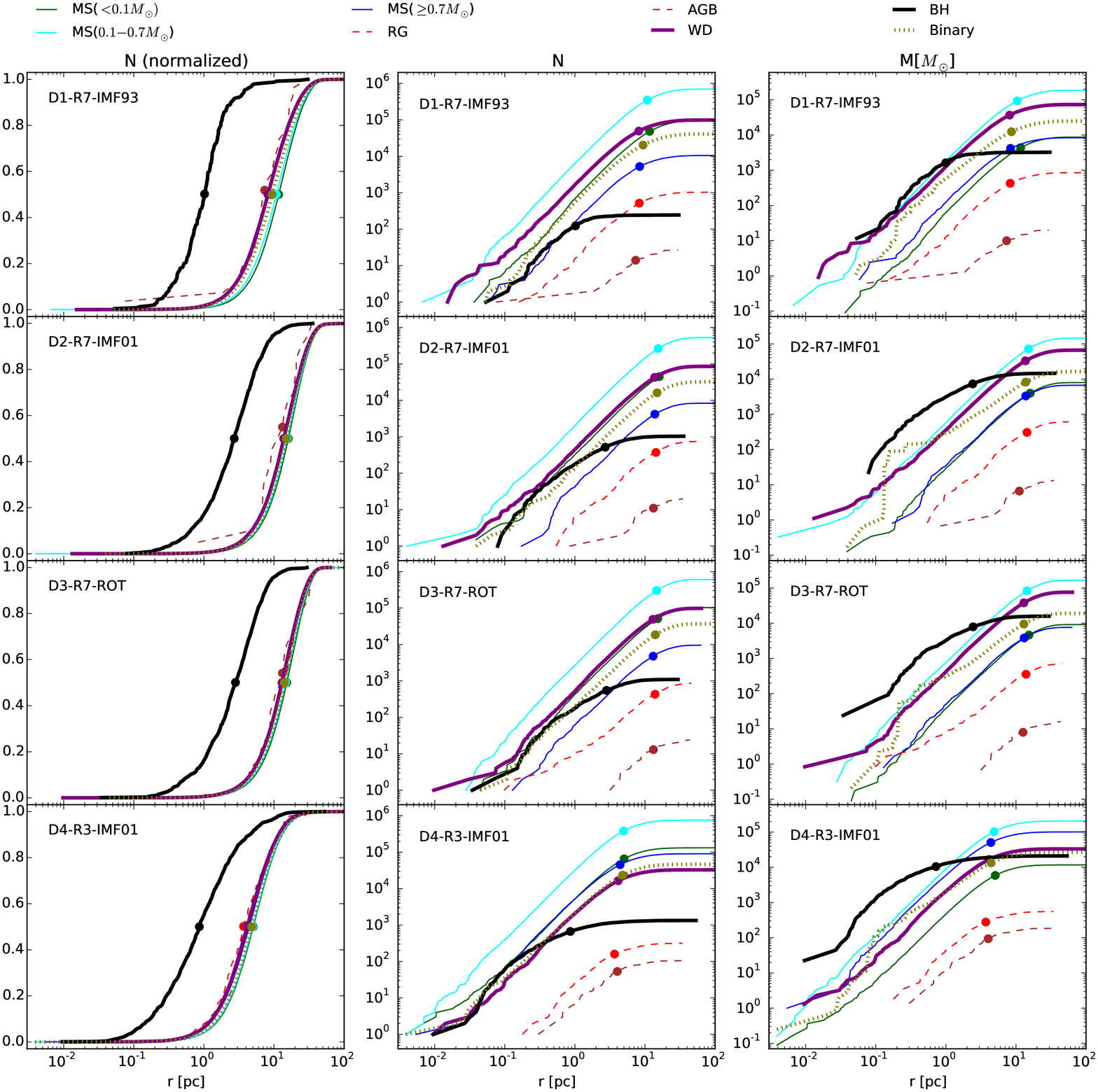} 
  \caption{Cumulative projected radial ($r$) distribution of different stellar types.
    R7 model data is obtained at $12$~Gyr and the D4-R3-IMF01 model uses $1$~Gyr.
    The left panels show the normalized cumulative number distribution $\bar{N}(r)$.
    The middle panels show the cumulative number distribution $N(r)$.
    The right panels show the cumulative mass distribution $M(r)$.
    The dot in the plots is the location of the radius enclosing half of the number (left and middle panels) and half of the mass inside (right panels).
  }
  \label{fig:cum}
\end{figure*}

To show the difference in the spatial distribution of the stellar components, we provide the cumulative surface number density profile ($N(r)$) and the cumulative surface mass profile ($M(r)$) of individual stellar components at $12$~Gyr in Fig.~\ref{fig:cum}. 
The normalized cumulative surface number density ($\bar{N}(r)$) shows that BHs are strongly mass segregated while all other components have no significant mass segregation.
The BHs of D1-R7-IMF93 are more centrally concentrated compared to D2-R7-IMF01 and D3-R7-ROT as also shown in Fig.~\ref{fig:snap}.
Although D4-R3-IMF01 only reaches $1$~Gyr, the BH subsystem has a similar level of mass segregation.
Only D1-R7-IMF93 shows a clear mass segregation feature as the WDs, AGB stars and RGs are more centrally concentrated as compared to the MSs with binaries in between.
There are no clear differences between the rotational (D3-R7-ROT) and non-rotational (D2-R7-IMF01) models.
In IMF01 (D2, D3 and D4) models, the total BH mass, $M_{\rm 2}$, is about $2.4 \times 10^4~M_{\rm \odot}$ and the combined mass of all other components $M_{\rm 1}$ is about $4.3 \times 10^5~M_{\rm \odot}$ around $100$~Myr.
The average mass of BHs, $m_{\rm 2}$, is about $16.7~M_{\rm \odot}$ and the other components have an average mass of $m_{\rm 1} \approx 0.46~M_{\rm \odot}$.
Due to the Spitzer mass segregation criterion $(M_{\rm 2}/M_{\rm 1})(m_{\rm 2}/m_{\rm 1})^{3/2}>0.16$ \citep{Spitzer1987}, the BHs in the IMF01 models are unstable systems and migrate quickly to the GC center and form a dense BH sub-cluster.
In the D1-R7-IMF93 model, the number of BHs is much smaller, but the mass segregation criterion can still be satisfied.
The long half-mass relaxation time $T_{\rm rh}$ in R7 models ($7-8$~Gyr initially, increasing to $>11$~Gyr after cluster expansion) is the reason for no clear segregation of other components.
The initial $T_{\rm rh}$ of D4-R3-IMF01 is about $1.9$~Gyr. 
Thus it also has no strong mass segregation at $1$~Gyr.
If the simulation continues, the segregation feature will become significant after $12$~Gyr.
Since the D1-R7-IMF93 model has a slightly shorter $T_{\rm rh}$ as compared to D2-R7-IMF01 and D3-R7-ROT, the mass segregation feature appears earlier.

The $N(r)$ without normalization shows one interesting phenomenon -- although BHs are more centrally concentrated, the innermost region ($0.02$-$0.07$~pc) of the R7 clusters have no BHs.
Instead, the WDs, MSs and binaries occupy the cluster density center (the mass density center of the cluster is treated as the center). 
The low-mass MSs ($< 0.7~M_\odot$) contribute most by number in the cluster.
This is due to the projection effect. 
When we inspect the three-dimensional profiles, this phenomenon is not noticeable.
It is expected that the innermost BHs do not exactly stay at the cluster center due to Brownian motions and the dynamical interaction involving BHs.
The $M(r)$ indicates that BHs are the most massive components within the central region.

\subsection{Structure evolution}
\label{sec:sp}

\begin{figure*}
  \centering
  \includegraphics[width=1.0\textwidth,height=!]{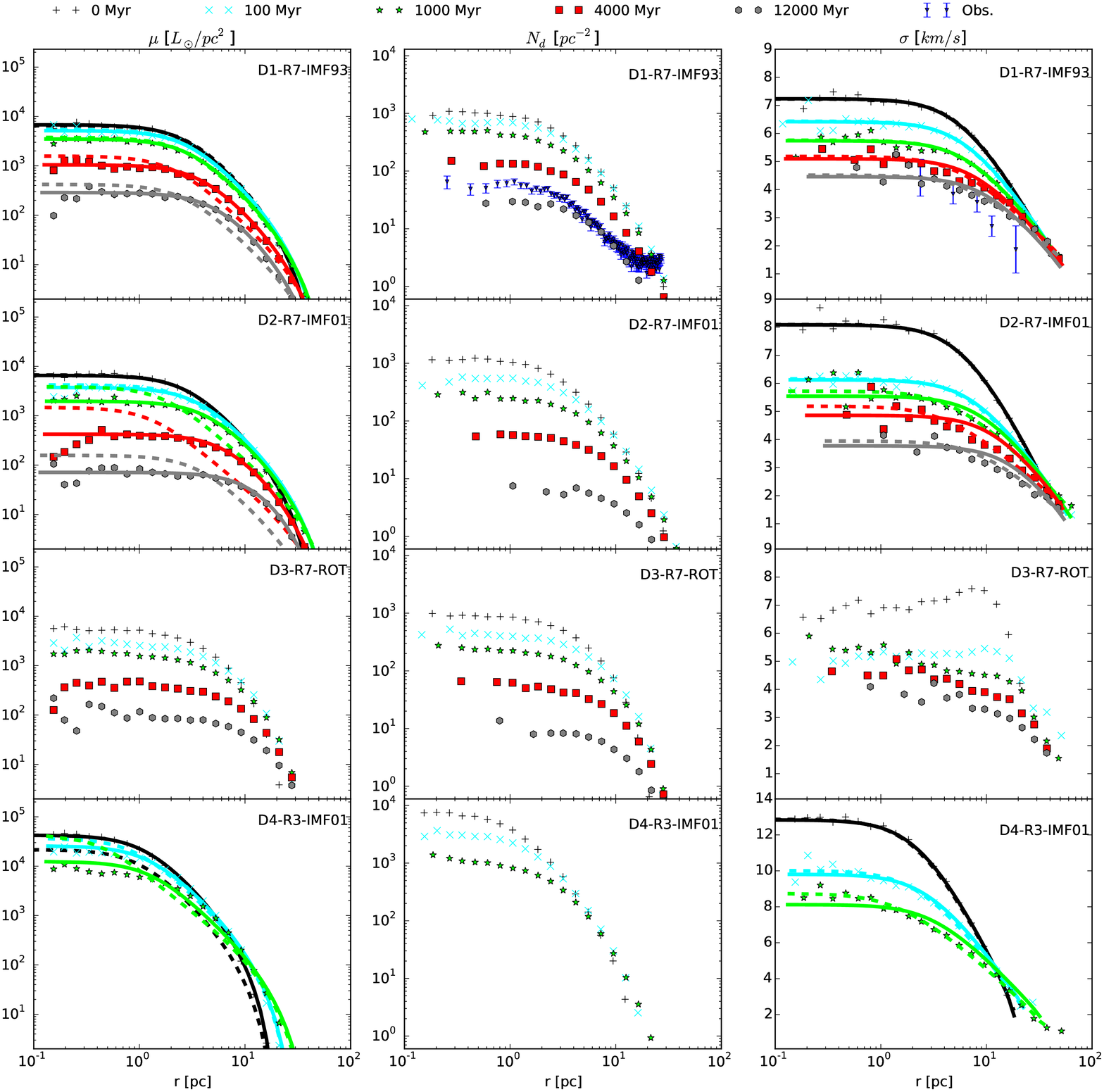} 
  \caption{$V$ band surface brightness profiles $\mu(r)$, surface number density profiles $N_{\rm d}(r)$ and line-of-sight surface velocity dispersion profile $\sigma(r)$ at different ages. 
    The markers are data from simulation. 
    To avoid the strong fluctuation generated by luminous stars like AGB stars, the $V$ band luminosity limit $20~L_\odot$ is used for $\mu(r)$.
    The observational data of $N_{\rm d}(r)$ and $\sigma(r)$ of NGC~4372 from Kacharov et al. (2014) are also shown in the upper panels.
    To be consistent with observations, the same $V$ band apparent magnitude upper limit $19$~mag (also the same distance modulus $(m-M)_V = 15.0$~mag without extinction correction) is used in the $N_{\rm d}(r)$ and $\sigma(r)$ simulation data.
    For each point in $N_{\rm d}(r)$ and $\sigma(r)$, the minimum number of stars is $50$ for better statistics.
    The King (1966) model fitting for $\mu(r)$ and $\sigma(r)$ with equal weights are shown as solid curves and with more weight of $\sigma(r)$ as dashed curves.
  }
  \label{fig:sp}
\end{figure*}

\subsubsection{Surface brightness profiles}
The observed surface brightness profile $\mu(r)$ of GCs provides information about the luminous mass distribution.
In the same way we show the mock $V$ band surface brightness profiles of five different ages for all models in Fig.~\ref{fig:sp} (left panels).
Typically a few bright RGs and AGB stars dominate the light at the center and generate strong fluctuations.
In Fig.~\ref{fig:sp} we have applied a $V$ band luminosity cutoff at $20~L_\odot$ (corresponding to $M_{\rm V}=1.58$~mag or $m_{\rm V}=15.39$~mag) to smoothen the profile \citep{Noyola2006,Chatterjee2013}.
When the clusters become old, their luminosity and surface brightness decrease.
This is driven by the evolution of massive stars into WDs, NSs and BHs, as well as the loss of stars in the tidal field of the central galaxy.
For a more top-heavy IMF (IMF01 models in Fig.~\ref{fig:sp}), the evolution is therefore stronger than in the IMF93 model.
In addition, the IMF01 models develop larger cores (the flat region of the $\mu(r)$) than the model with a low fraction of massive stars (IMF93).

\subsubsection{Surface number density profile}

The 'observed' stellar surface number density profile $N_{\rm d}(r)$ is shown in the middle panels of Fig.~\ref{fig:sp}. 
While the $\mu(r)$ can be obtained directly from adding the stellar fluxes, the number density estimate requires the identification of individual bright stars.
We assume a luminosity limit of $L_{\rm V} > 2.15~L_\odot$, corresponding to $m_{\rm v} \sim 19$~mag with a distance modulus $(m-V)_V = 15$~mag.
A similar assumption was made to construct the surface number density distribution of NGC~4372 \citep{Kacharov2014}.
Most stars are not accounted for, as they are faint MS stars falling below the luminosity (detection) limit.
Therefore, the number density estimates suffer from large systematic and statistical uncertainties.
Despite these limitations, the overall distribution of bright stars resembles the surface brightness distribution (not too surprisingly as they dominate the flux).

\subsubsection{Line-of-sight velocity dispersion profile}
\label{sec:vdp}

The line-of-sight velocity dispersion profiles $\sigma(r)$ provide information on the cluster dynamics and are shown in the right panels of Fig.~\ref{fig:sp}, assuming the same luminosity cutoff as for $N_{\rm d}(r)$.
As compared to $\mu(r)$ and $N_d(r)$, $\sigma(r)$ shows smaller cores at $12$~Gyr.
The D2-R7-IMF01 model has initially a $1$~km/s higher central velocity dispersion $\sigma(0)$ as compared to D1-R7-IMF93, but $\sigma(0)$ decreases faster and becomes $0.5$~km/s smaller at $12$~Gyr.
The rotational model D3-R7-ROT shows very different $\sigma(r)$ compared to D2-R7-IMF01 during the first $100$~Myr.
But this rotational feature does not last long since the discrepancy of these two models is not significant after $1$~Gyr.
At $12$~Gyr, they have a very similar $\sigma(r)$.

\subsubsection{King model fitting}

\begin{table}
  \centering
  \caption{King model fitting parameters at different stages. $W_0$ is scaled central potential; $c$ is concentration; $M$ is cluster total mass.
The suffix ``e'' means K66-E fitting with equal weights of $\mu(r)$ and $\sigma(r)$ and the suffix ``v'' means K66-V fitting with more weight of $\sigma(r)$. Others (no suffix) are data obtained directly from simulations.
The unit of Time is Gyr and unit of $M$ is $10^5~M_\odot$.
}
  \label{tab:king}
  \begin{tabular}{l|l cc cc ccc cc cc}
\hline\hline
 & Time & $W_{0,e}$ & $W_{0,v}$ & $c_{e}$ & $c_{v}$ & $M_{e}$   & $M_{v}$    & $M$   \\\hline
& 0 & 6.08 & 5.89 & 1.27 & 1.23  & 4.77 & 4.83 & 4.75 \\
& 0.05 & 6.12 & 6.38 & 1.28 & 1.35  & 4.46 & 4.28 & 4.36 \\
& 0.1 & 6.24 & 6.30 & 1.31 & 1.33  & 4.25 & 4.18 & 4.23 \\
& 0.5 & 6.29 & 6.17 & 1.33 & 1.30  & 3.84 & 3.88 & 3.91 \\
& 1 & 6.34 & 6.50 & 1.34 & 1.38  & 3.69 & 3.60 & 3.77 \\
& 2 & 6.50 & 6.70 & 1.38 & 1.44  & 3.38 & 3.36 & 3.58 \\
& 4 & 5.98 & 7.21 & 1.25 & 1.59  & 3.31 & 2.98 & 3.37 \\
& 6 & 5.65 & 6.61 & 1.17 & 1.41  & 3.12 & 2.94 & 3.23 \\
& 8 & 5.45 & 6.23 & 1.12 & 1.31  & 2.99 & 2.90 & 3.13 \\
& 10 & 5.50 & 6.44 & 1.13 & 1.37  & 2.87 & 2.78 & 3.04 \\
\multirow{-11}{*}{\rotatebox{90}{D1-R7-IMF93}}& 12 & 5.31 & 6.97 & 1.09 & 1.52  & 2.83 & 2.60 & 2.96 \\
\hline
& 0 & 5.97 & 6.00 & 1.25 & 1.26  & 6.15 & 6.09 & 5.92 \\
& 0.05 & 6.12 & 6.29 & 1.28 & 1.33  & 4.80 & 4.72 & 4.81 \\
& 0.1 & 6.15 & 7.21 & 1.29 & 1.39  & 4.66 & 4.47 & 4.58 \\
& 0.5 & 6.17 & 6.17 & 1.30 & 1.59  & 4.43 & 3.90 & 4.12 \\
& 1 & 5.89 & 7.61 & 1.23 & 1.72  & 4.40 & 3.63 & 3.92 \\
& 2 & 5.72 & 7.45 & 1.19 & 1.67  & 4.06 & 3.27 & 3.66 \\
& 4 & 5.08 & 8.03 & 1.04 & 1.84  & 4.02 & 2.82 & 3.34 \\
& 6 & 4.53 & 7.01 & 0.94 & 1.53  & 3.62 & 2.60 & 3.11 \\
& 8 & 4.03 & 7.48 & 0.84 & 1.67  & 3.45 & 2.37 & 2.90 \\
& 10 & 3.55 & 6.33 & 0.76 & 1.34  & 3.10 & 2.30 & 2.71 \\
\multirow{-11}{*}{\rotatebox{90}{D2-R7-IMF01}}& 12 & 3.30 & 6.47 & 0.72 & 1.38  & 3.03 & 2.22 & 2.53 \\
\hline
& 0 & 6.00 & 6.16 & 1.25 & 1.29  & 6.08 & 5.88 & 5.92 \\
& 0.05 & 6.50 & 6.58 & 1.38 & 1.41  & 4.83 & 4.60 & 4.81 \\
& 0.1 & 6.46 & 6.90 & 1.37 & 1.50  & 4.61 & 4.22 & 4.58 \\
& 0.5 & 7.39 & 7.71 & 1.65 & 1.75  & 4.03 & 3.58 & 4.13 \\
\multirow{-5}{*}{\rotatebox{90}{D4-R3-IMF01}}& 1 & 6.94 & 8.39 & 1.54 & 1.95  & 4.00 & 3.16 & 3.95 \\
\hline\hline
  \end{tabular}
\end{table}

There are two widely used methods to extract structural parameters of GCs, such as the core radius $R_{cl}$ and the half-light radius $R_{hl}$.
The first method is using a \cite{King1962} profile to fit either $\mu(r)$ or the cumulative surface brightness profile.
\cite{Morscher2015} suggest that the latter can provide better fits.
Thus we use Eq.~18 in \cite{King1962} (it includes the tidal cutoff $R_{\rm tl}$) to fit the cumulative $V$ band surface brightness profile.
The fitting parameters $R_{\rm cl}$ and $R_{\rm tl}$ can then be used to reconstruct the surface brightness profile.

The \cite{King1962} profile fitting (hereafter K62) is easy to handle and commonly used to analyze observations of GCs.
It provides good estimates of $R_{\rm hl}$ and $R_{\rm cl}$.
However, K62 uses an empirical formula and the parameters $R_{\rm hl}$ and $R_{\rm cl}$ give two-dimensional information.
Thus a more sophisticated method (but more complicated) is to use the \cite{King1966} single-mass, isotropic model (hereafter K66) to fit both the $\mu(r)$ and the velocity dispersion profiles $\sigma(r)$.
Thus the three-dimensional core radii can be obtained.
However, the different weights of $\mu(r)$ and $\sigma(r)$ give very different results.
Thus the fitting with equal weights of two profiles (K66-E; solid curves) and more weight of $\sigma(r)$ (K66-V; dashed curves) are shown together in Fig.~\ref{fig:sp} for comparison.
The K62 fitting curves of $\mu(r)$ are similar to the results of K66-E; they are not shown in Fig.~\ref{fig:sp} to avoid confusion.
For the K66-E fitting, the goodness of fit value was calculated using an equal contribution from the individual functions which fit the $\mu(r)$ and $\sigma(r)$. 
In the case of the K66-V fitting, the contribution for fitting the $\sigma(r)$ was increased by a factor $50$.
This value was selected based on the experiments for obtaining the best fitting.
For most of the snapshots, the number of data points in $\sigma(r)$ were fewer than those in $\mu(r)$, so a large value was needed to increase the contribution of $\sigma(r)$.
The $\mu(r)$ bins at very extended radii in which the surface brightness was too faint with large fluctuations (below $2.3\times 10^3 L_{\rm \odot/arcsec^{2}}$) were ignored. 
Similarly, few outer bins at extended radii where a flattening and increase in the velocity dispersion \citep{Lane2009,Kupper2010} is observed due to the Galactic tidal field were also ignored in order to improve the fitting.
For $\mu(r)$, the K62 and K66-E fitting methods provide almost the same results since the solid and dashed curves overlap each other in most cases (there is difference at $1$~Gyr in D4-R3-IMF01).
They are also consistent with the simulation data.

However, the K66-E fitting curves of $\sigma(r)$ are much worse than $\mu(r)$.
This is more pronounced in IMF01 models.
In contrast, K66-V fits $\sigma(r)$ very well, but not $\mu(r)$.
This indicates that the \cite{King1966} model is not fully consistent with our GC models for surface brightness distribution and dynamics.

The K66 fitting can provide useful structural parameters of GCs, including parameters such as the scaled central potential $W_{\rm 0}$, the concentration $c$, the three-dimensional core radius $R_{\rm ck}$ and the total cluster mass $M$.
These fitting parameters (both K66-E and K66-V; $R_{\rm ck}$ is shown later) are listed in Table~\ref{tab:king}. 
The $\sigma(r)$ of the rotational model D3-R7-ROT cannot be fitted by K66, thus D3-R7-ROT is not listed.
The total mass $M$, directly obtained from the simulation data, is shown for comparison.
The K66 model has only one free parameter, $W_0$. 
The concentration $c=\log(R_{\rm t}/R_{\rm ck})$ gives the ratio of tidal radius and core radius.
When $W_0$ increases, $c$ also increases.
Usually when a cluster is in the tidally filling stage, its tidal radius is determined by its total mass and by the host environment. 
For a fixed tidal radius, a larger $c$ means a smaller core radius.
The initial $W_{\rm 0}$, $c$ and $M$ are the same and consistent with our models ($W_0=6$ and $c=1.255$).
Then the divergence between K66-E and K66-V grows significantly after $4$~Gyr for R7 models.
D2-R7-IMF01 has $W_{\rm 0,v}$ ($c_{\rm v}$) almost twice of $W_{\rm 0,e}$ ($c_{\rm e}$) after $12$~Gyr.
The value of $c_{\rm v}$ after $12$~Gyr increases by $20\%$ since the beginning, which indicates the core is shrinking relative to $R_{\rm t}$.
But after $12$~Gyr, $c_{\rm e}$ decreases significantly by $42\%$, which suggests that the core is expanding relatively fast.
In the D1-R7-IMF93, a similar feature occurs but not as strong as for D2-R7-IMF01.
The difference between $c_{\rm e}$ and $c_{\rm v}$ grows faster in D4-R3-IMF01 since after $1$~Gyr it is already significant.
This is probably due to its short relaxation time.

These two fitting curves provide completely opposite results for the cluster evolution.
As discussed above, K66-E is consistent with $\mu(r)$ and K66-V is consistent with $\sigma(r)$.
The $\mu(r)$ represents the luminosity density distribution which only includes the luminous stars, while $\sigma(r)$ is affected by all the mass in the clusters including ``dark'' objects (BHs).
Thus the divergence of fitting suggests that there are two cores in the clusters.
Below we will provide a more detailed discussion about these two core structures.

The dynamical mass estimated from K66 fitting is also different from the real $M$ after $12$~Gyr.
In D1-R7-IMF93, both $M_{\rm e}$ and $M_{\rm v}$ are underestimated, and $M_{\rm e}$ is closest to $M$.
But in D2-R7-IMF01 and D4-R3-IMF01, $M_{\rm e}$ is overestimated while $M_{\rm v}$ is underestimated.
These different behaviors in IMF93 and IMF01 models and the departure from the real value of $M$ may be caused by the different structures of the clusters.

\subsubsection{Comparison with NGC~4372}

Since we use NGC~4372 as a reference, it is interesting to check whether our models have a similar observational structure.
Unfortunately, the only available $\mu(r)$ of NGC~4372 is from \cite{Trager1995} and has too poor quality for a comparison at the level we require.
Instead, \cite{Kacharov2014} provide good surface number density profiles (hereafter $N_{\rm d,K}(r)$) and line-of-sight velocity dispersion profiles of RGs and AGB stars (hereafter $\sigma_{\rm k}(r)$).
Since the D1-R7-IMF93 model has the most similar half-light radius as NGC~4372, we show the observational data together with the D1-R7-IMF93 data in Fig.~\ref{fig:sp}.
As mentioned above, our $N_{\rm d}(r)$ has the same magnitude cutoff as $N_{\rm d,K}(r)$.
The $N_{\rm d}(r)$ at $12$~Gyr overlaps with the observed $N_{\rm d,K}(r)$ in the region from $5$ to $10$~pc.
Below $5$~pc, $N_{\rm d,K}(r) \approx 2 \times N_{\rm d}(r)$.
Since D1-R7-IMF93 has about twice the $R_{\rm cl}$ of NGC~4372, it is expected to see a larger core in the D1-R7-IMF93 model at $12$~Gyr.
Above $10$~pc, $N_{\rm d,K}(r)$ becomes flat while $N_{\rm d}(r)$ continues the decreasing trend.
The tidal radius of D1-R7-IMF93 at $12$~Gyr is about $52$~pc, which is far from the flat transformation radius $20$~pc as seen in the observational data.
One possibility is that the field star contamination contributes to the flat part since $N_{\rm d,K}(r)$ decreases to $2$-$4$~pc$^{-2}$ beyond $20$~pc.
If this is true, the higher central $N_{\rm d,K}(r)$ might also be caused by contamination.

For the $\sigma_{\rm K}(r)$, since it only includes about one hundred RGs and AGB stars, there are only five data points with a very large uncertainty.
The $\sigma_{\rm K}(r)$ seems steeper than $\sigma(r)$ for D1-R7-IMF93 at $12$~Gyr.
We also try to use only RGs and AGB stars to obtain a similar $\sigma(r)$ as $\sigma_{\rm K}(r)$.
The new $\sigma(r)$ shows no difference as compared to the original $\sigma(r)$.

\subsection{Two- and three-dimensional profiles}

\begin{figure}
  \centering
  \includegraphics[width=0.5\textwidth,height=!]{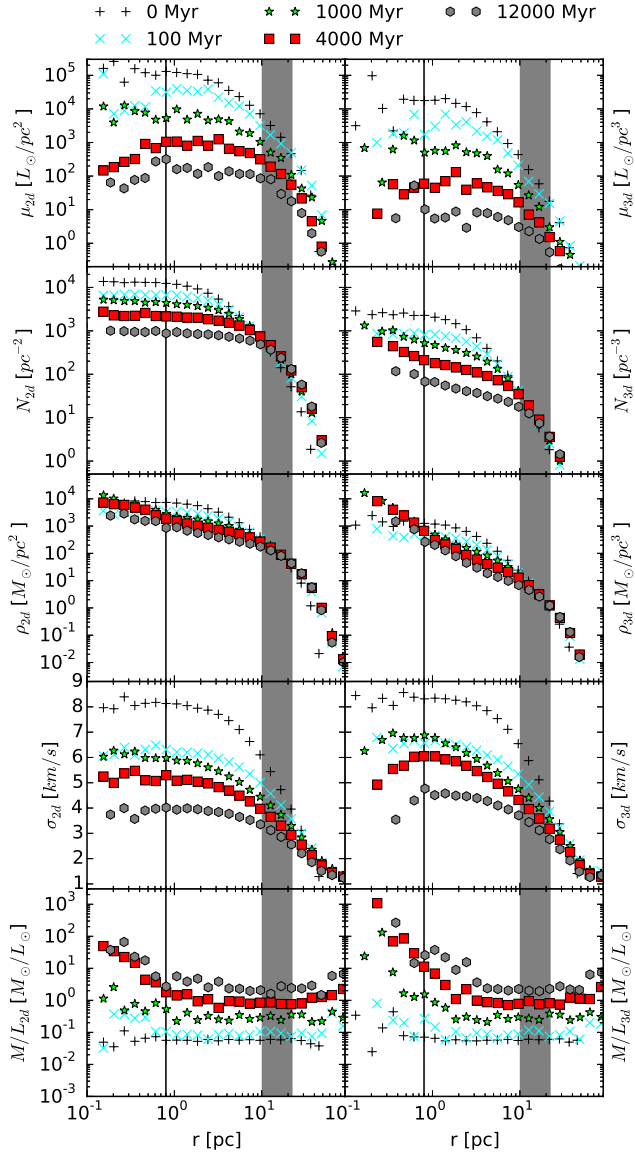}
  \caption{Two- and three-dimensional profiles of R7~IMF01.
    The left panels show two-dimensional surface profiles (projected on a plane) and the right panels show three-dimensional radial profiles (spherical shells).
    $\mu_{\rm 2d}(r)$: $V$ band surface brightness profiles; $\rho_{\rm 2d}(r)$: surface density profile; $\sigma_{\rm 2d}(r)$: surface line-of-sight velocity dispersion profile; $M/L_{\rm 2d}(r)$: surface mass-to-light ($V$ band) ratio profile;
    $\mu_{\rm 3d}(R)$: $V$ band radial brightness profiles; $\rho_{\rm 3d}(R)$: radial density profile; $\sigma_{\rm 3d}(R)$: radial one-dimensional velocity dispersion profile; $M/L_{\rm 3d}(R)$: radial mass-to-light ratio profile.
    All objects in the clusters are counted. 
    The black vertical line shows the approximate inner core radius and the outer core radius is within the grey region.
 }
  \label{fig:spv}
\end{figure}

The two-dimensional profiles with magnitude cutoff in Fig.~\ref{fig:sp} show the typical features of GCs that are generally observed in the Milky Way.
In Fig.~\ref{fig:spv} we show two-dimensional profiles without a magnitude cutoff, together with the three-dimensional distributions of the D2-R7-IMF01 model.
As compared to the $\mu(r)$ in Fig.~\ref{fig:sp}, $\mu_{\rm 2d}(r)$ in Fig.~\ref{fig:spv} has much higher initial central brightness.
After $12$~Gyr, they are similar.
There is also no structure difference between projected $\mu_{\rm 2d}(r)$ and spherical $\mu_{\rm 3d}(R)$.

The $N_{\rm 2d}(r)$ is much smoother compared to the $\mu_{\rm 2d}(r)$ in the core.
However, the $N_{\rm 3d}(R)$ has an increasing trend towards the center after $1$~Gyr. 
This increasing feature is hidden by the projection effect in the two-dimensional $N_{\rm 2d}(r)$.
The surface mass density profile $\rho_{\rm 2d}(r)$ and radial density profile $\rho_{\rm 3d}(R)$ are also shown together.
Similar to $N_{\rm 3d}(R)$, the $\rho_{\rm 2d}(r)$ shows a two-core structure which is more pronounced in $\rho_{\rm 3d}(R)$ (the inner core is more like a cusp).
This feature is consistent with our finding in the K66-E and K66-V fitting discussed above.
The size of inner core in $\rho_{\rm 2d}(r)$ is $\sim 1$~pc and the outer core is $10$-$20$~pc.
Thus the inner core is consistent with the K66-V fitting ($\sigma(r)$) and the outer core is related to the K66-E fitting ($\mu(r)$).
The inner core feature is more significant in the mass density profile and disappears in the surface brightness profile.
This means that it is populated by the dark objects (BHs). 
As shown in Fig.~\ref{fig:cum}, the BH subsystem is centrally concentrated and its projected half-mass radius is $2$-$3$~pc at $12$-Gyr, which is very close to the inner core size.
This also explains why there is a difference between the K66-E and K66-V fittings.
Since the D2-R7-IMF01 model has many more BHs than D1-D7-IMF93, this feature is stronger in D2-R7-IMF01.

The $\sigma_{\rm 2d}(r)$ is identical to $\sigma(r)$ but the radial one-dimensional velocity dispersion profile $\sigma_{\rm 3d}(R)$ decreases in the inner core at $1$~Gyr. 
The minimum number of stars per bin for velocity dispersion calculation is $50$, thus this feature should not be influenced too much by stochastic effects.
Since BHs have much larger mass and the velocity dispersion should be lower due to energy equipartition (mass segregation), a lower $\sigma_{\rm 3d}(R)$ in the inner core is expected.
In the projected $\sigma_{\rm 2d}(r)$, the central value is smoothed by the stars distributed along the line-of-sight.

The mass-to-light ratio $M/L_{\rm 2d}(r)$ and $M/L_{\rm 3d}(R)$ are also shown together.
It is clear that due to the presence of the BH subsystem both the projected and the spherical $M/L$ are significant larger in the inner core.
The outer regions of clusters show no variation of $M/L$ along the radial direction.

\subsection{Radius evolution}
\label{sec:radius}

\begin{figure}
  \includegraphics[trim=0 2cm 0 0, width=0.5\textwidth,height=!]{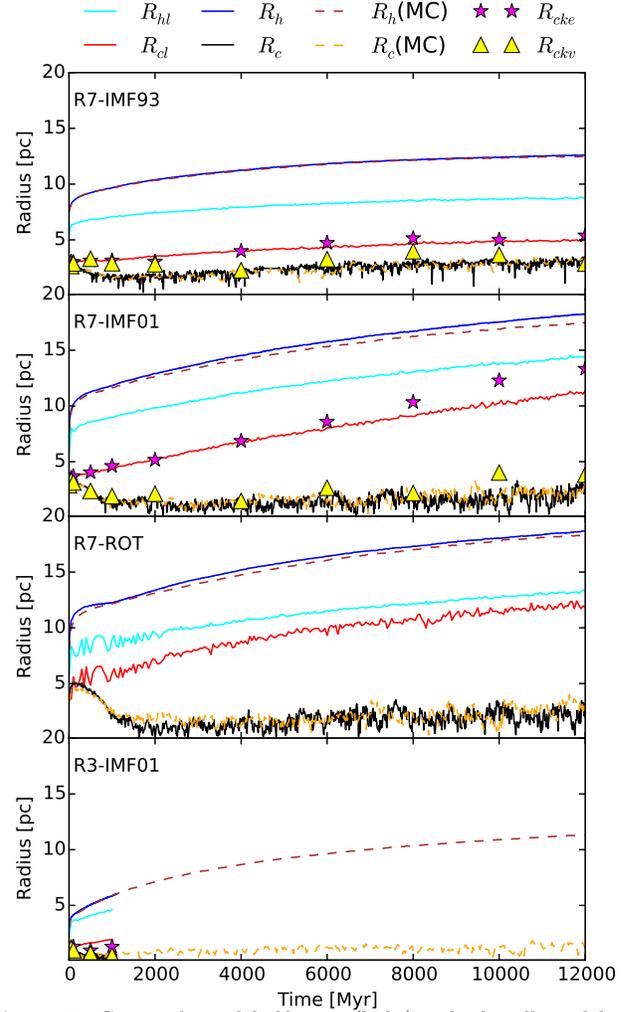}
  \caption{Core radii and half-mass (light) radii for all models. 
    $R_{\rm c}$ is the three-dimensional core radius calculated by the generalized mass-square weighted method from Casertano \& Hut (1985).
    $R_{\rm h}$ is three dimensional half-mass radius.
    $R_{\rm cl}$ and $R_{\rm hl}$ are projected core radius and half-light radius obtained from King (1962) profile fitting. 
    $R_{\rm ck}$ is the three-dimensional core radius obtained from King (1966) model fitting.
    The dashed curves are data from MOCCA simulations with the same definitions of $R_{\rm c}$ and $R_{\rm h}$.
    }
  \label{fig:radius}
\end{figure}

Fig.~\ref{fig:snap} indicates very different sizes of GCs at $12$~Gyr as a result of the different IMFs, and Section~\ref{sec:sp} shows a very different behavior of the cores in the surface brightness and velocity dispersion profiles.
In Fig.~\ref{fig:radius} we provide the radii evolution of all models.
Before discussing the results, we first clarify the different definitions of radii in the figure.
The K62 fitting provides the projected core radius $R_{\rm cl}$, and K66 fitting gives the three-dimensional core radius, in which $R_{\rm cke}$ is from K66-E fitting and $R_{\rm ckv}$ is from K66-V (see Section~\ref{sec:sp}).
However, in theoretical modeling of star clusters, the core radius has different definition.
In the NBODY6++GPU code, we use the generalized mass-square weighted method from \cite{Casertano1985} to calculate the three-dimensional core radius (hereafter $R_{\rm c}$), which take into account both luminous and dark objects (BHs).
Fig.~\ref{fig:radius} indicates that $R_{\rm c}$ and $R_{\rm ckv}$ are consistent and $R_{\rm cl}$ and $R_{\rm cke}$ are similar.
It means the theoretical core radius can be well recovered by K66 fitting of the velocity dispersion profile and $R_{\rm cl}$ and $R_{\rm cke}$ represent the luminous stellar core observed in the surface brightness profile.
There is a small difference between $R_{\rm cl}$ and $R_{\rm cke}$. 
This is due to the projection effect.

It is also evident that the $R_{\rm c}$ or $R_{\rm ckv}$ shows core collapse features in all models ($R_{\rm ckv}$ in D1-R7-IMF93 does not match $R_{\rm c}$ very well).
The core collapse is terminated after $1$~Gyr in R7 models and after $200$~Myr in D4-R3-IMF01.
After collapse, the $R_{\rm c}$ is stable at $2$-$3$~pc with a slightly increasing trend until $12$~Gyr (R7 models).
There is also the core oscillation feature discussed in previous studies (e.g. \citealp{Heggie2009,Hurley2012}).
However, for all models, $R_{\rm cl}$ ($R_{\rm cke}$) increases all the time.
Thus the core collapse shown in $R_{\rm c}$ ($R_{\rm ckv}$) is dominated by the BH subsystem, which is different from the classical core-collapse phases for all populations.
Since IMF01 models have larger BH subsystems, the divergence between $R_{\rm cl}$ ($R_{\rm ck}$) and $R_{\rm c}$ is more pronounced. 
This is also consistent with the difference of $W_{\rm 0}$ and $c$ evolution discussed in Section~\ref{sec:sp}.
After $12$~Gyr, in D2-R7-IMF01 $R_{\rm cke}$ is even larger than $11$~pc with a ratio of $R_{\rm cke}/R_{\rm c} \approx 3.7$.

With the K62 fitting, the two-dimensional half-light radius $R_{\rm hl}$ is also easy to calculate.
The $R_{\rm hl}$ in D1-R7-IMF93 and D2-R7-IMF01 are both about $4$~pc smaller than $R_{\rm h}$.
This difference is mostly built in the early expansion phase of clusters due to stellar mass loss.
After that, $R_{\rm hl}$ and $R_{\rm h}$ increase in a parallel trend.
But in the rotational model D3-R7-ROT, this difference grows all the time.
In D3-R7-ROT, the $R_{\rm cl}$ tends to be closer to $R_{\rm hl}$ and will probably overlap for a few Gyr after $12$~Gyr.

The different IMFs also result in very different expansion in the early stage (a few hundred Myr) of clusters.
The IMF01 models have larger stellar mass loss due to more massive stars.
The $R_{\rm h}$ in D2-R7-IMF01 expanded up to $12$~pc after $200$~Myr while in D1-R7-IMF93 it is about $9$~pc.

The $R_{\rm h}$ and $R_{\rm c}$ obtained from MOCCA simulation data are also provided for comparison.
The MOCCA data in Fig.\ref{fig:radius} indicates that the simulations show very good agreement with the direct $N$-body simulations for both $R_{\rm h}$ and $R_{\rm c}$. 
This further strengthens the results presented in \cite{Giersz2013} which claims that the MOCCA code follows closely the $N$-body simulations for different global cluster properties and for different environments.
Besides, MOCCA also reasonably closely follows the evolution of the rotational model (D3-R7-ROT), which is a big surprise because the code is not designed to follow non-spherically symmetric systems.

\cite{Breen2013} derived a relation for BH half-mass radius $R_{\rm h,bh}$ and the cluster's $R_{\rm h}$ relation in the energy balance phase between the BH subsystem and the cluster for two-mass component systems (Eq.~4).
Our data shows that their model is consistent with the realistic simulations of GCs (Fig.~2 in Breen \& Heggie).
In our results, the D2-R7-IMF01 has a $R_{\rm h,bh}/R_{\rm h} \approx 0.175$ at $12$~Gyr and it did not change significantly after $4$~Gyr ($R_{\rm h,bh}/R_{\rm h} \approx 0.17$ at $4$~Gyr).
It indicates that the BH subsystem already reaches the balance phase at $4$~Gyr.
The corresponding value of $M_{\rm 2}/M_{\rm 1}$ is $0.061$ and $m_{\rm 2}/m_{\rm 1}$ is $41$ at $12$~Gyr.
Here $M_{\rm 2}$ ($M_{\rm 1}$) is the total mass of BHs (other objects) and $m_{\rm 2}$ ($m_{\rm 1}$) is the average mass of BHs (other objects).
Compared with Fig.~2 in Breen \& Heggie, the ratio $R_{\rm h,bh}/R_{\rm h}$ of D2-R7-IMF01 is slightly lower than their result ($\sim 0.2$), which means the BHs are slightly more centrally concentrated.
For D1-R7-IMF93, $M_2/M_1 \approx 0.011$, $m_2/m_1 \approx 40$ and $R_{\rm h,bh}/R_{\rm h} \approx 0.098$ at $12$~Gyr, which is also slightly lower than the Breen \& Heggie result ($\sim 0.12$).
This difference can be caused by the multiple-mass system instead of two-mass components, the large $N$ or the stellar evolution during the first few hundred Myr.
In rotational model D3-R7-ROT, $M_2/M_1 \approx 0.058$, $m_2/m_1 \approx 42$, $R_{\rm h,bh}/R_{\rm h} \approx 0.19$ at $12$~Gyr, which is similar to the result of Breen \& Heggie.
This suggests the cluster rotation has no strong influence on the energy balance between the BH subsystem and the cluster.

Compared with NGC~4372, the D1-R7-IMF93 model has a larger size $R_{\rm hl}$ ($8.7$~pc) and $R_{\rm cl}$ ($5.0$~pc) at $12$~Gyr.
The $R_{\rm hl}$ is $0.3$ times larger and $R_{\rm cl}$ is $0.7$ times larger than in NGC~4372.

\section{Conclusion}
\label{sec:con}

In this paper we present the DRAGON GC models which are evolved from the initial phase after gas removal up to $12$~Gyr.
The DRAGON clusters are four realistic direct $N$-body simulations of GCs with initially one million stars, including $5\%$ primordial binaries, both single and binary stellar evolution and tidal fields together for the first time.
It is a very computational intensive task to simulate even one such model, so it cannot always replace more approximate methods based on the Fokker-Planck approximation (as in MOCCA; e.g. \citealp{Leigh2013,Leigh2015,Giersz2013,Giersz2015}). 
However, our models can serve as prototypes of GCs, to be compared with approximate models (e.g. the comparison with half-mass radius and core radius of MOCCA models in Fig.~\ref{fig:radius}) and also to gain some insight into the detailed physical processes which play a role in the evolution of clusters.
On the other hand, we have full information about the modeled cluster, regarding its population and color gradients, its projected stellar densities and kinematic properties of the luminous and ``dark'' component.
``Dark'' here refers to the low-luminosity stellar objects or remnants, which are too faint to be observed, and the dark objects like BHs.
Such data can then be compared with results of the most advanced approximate methods (like the MOCCA code) to assess their reliability.

Subsequently, we presented ``simulated observations'' of our models and compared some of these data with the reference cluster NGC~4372.
Due to the more fundamental goals of this work we do not aim to reproduce precisely all features of NGC~4372. 
For example, we do not model the time varying tidal fields resulting from its eccentric orbit. 
Instead, we adopt a steady tidal field as is typical for a cluster on a circular orbit, as is used in many previous cluster simulations and is standard in many codes (e.g. \citealp{Heggie2014}). 
Also, in this paper, we have not yet studied exotic objects that originate from tidally interacting binaries, or relativistic interactions between components of binaries with compact remnants.
All these processes are already included in the code, but will be the subject of a future detailed study.
For these reasons we prefer to name our simulated clusters as ``DRAGON'' clusters, rather than ``a model for NGC~4372''. 
In this paper we have focused on the global dynamics, the structure and kinematics and the distinction between luminous and non-luminous objects over the entire lifetime of the cluster. 
More precisely, with use of the COCOA code \citep{Askar2014} we performed the data analysis in the observational way - we transformed our data into the photometric images observed by ``customized'' telescopes, constructed the color-magnitude diagram, spatial distribution of different stellar components, and analyzed the surface brightness profiles and line-of-sight velocity dispersion profiles.
Then we can compare the observed parameters with our models to check whether the standard observational data analysis can represent the real data obtained directly from the simulations.

Our main finding is that the presence of a large number of stellar-mass BHs as a non-luminous subsystem in the core of the cluster produces a relatively large luminous stellar core radius and half-mass radius.
This has already been reported by \cite{Merritt2004,Hurley2007,Mackey2008,Banerjee2010,Downing2012,Breen2013,Morscher2015,Giersz2015}.
But all previous papers used some approximate models, smaller particles numbers, analytic or semi-analytic methods.
Our result for the first time confirms this core of stellar-mass BHs in full direct $N$-body simulations of large particle numbers.
The previous studies suggest that only a few or zero BHs can be retained in GCs after $12$~Gyr.
However, a two-mass semi-analytic model from \cite{Breen2013} indicates that a BH subsystem can survive a very long time in GCs with four evolutionary phases.
In our case, BHs finish the mass segregation (core collapse) after $1$~Gyr for R7 models and remain in the balanced evolution phase until $12$~Gyr. 
The situation is similar for D4-R3-IMF01, but with a shorter timescale ($\sim 200$~Myr).
The comparison of $R_{\rm h,bh}/R_{\rm h}$ in our models is consistent with the Breen \& Heggie result with a slightly more concentrated BH subsystem.

The presence of the subsystem of stellar-mass BHs is so pronounced that we can even see a double core structure (see the mass density profiles in Fig.~\ref{fig:spv}) - the inner core formed by the black hole components, and an outer core formed by the observable stars.
From the evolution of different core radii shown in Fig.~\ref{fig:radius}, we can see that the theoretical core radius $R_{\rm c}$ and core radius obtained from \cite{King1966} model fitting to the velocity dispersion profile ($R_{\rm ckv}$) show core collapse feature (inner core) while the luminous stellar core radius $R_{cl}$ obtained from surface brightness profiles continue expansion during the whole evolution (outer core).
This opposite trend of core evolution is driven by the BH subsystem. 
It also suggests that an inconsistent result from \cite{King1966} fitting to $\mu(r)$ and $\sigma(r)$ can be a potential method to identify BH subsystems in star clusters for observers.
However, this requires very good spectroscope observations of individual stars at the cluster center, which is usually very challenging.

We also find that relatively small differences in the IMF can significantly affect the dynamical evolution of GCs.
Fig.~\ref{fig:snap} shows completely different densities of D1-R7-IMF93  and D2-R7-IMF01.
The D2-R7-IMF01 model has much lower density than D1-R7-IMF93.
Since IMF01 is more top-heavy, the mass loss driven by stellar evolution is much stronger during the first $100$~Myr.
Thus the difference is already achieved during the early stages (see Fig.~\ref{fig:radius}).
Subsequently, the much larger number of BHs in D2-R7-IMF01 drives the cluster towards a faster expansion phase.

\section{Discussion}
\label{sec:dis}

Surprisingly, our models suggest that initial rotation has no major impact on the late evolution of the clusters, which differs from claims made for Fokker-Planck models and some $N$-body simulations (but with smaller $N$), such as those of \cite{Kim2008} and \cite{Hong2013}. 
We agree with the Fokker-Planck models in one important aspect, a point made already by \cite{Einsel1999}, namely that the cluster rotation monotonously decreases over time through the combined effect of two-body relaxation, transport of angular momentum outwards and mass loss due to the tidal field. 
However, in our model D3-R7-ROT we cannot confirm that the evolution with rotation is much faster than D2-R7-IMF01. 
This should be the subject for further study, it could again be due to the different setup of the initial models or stellar evolution.

The single-mass, isotropic assumption in \cite{King1966} is a poor approximation for GCs with BH subsystems.
Thus our models have an inconsistent King fitting to surface brightness profiles and velocity dispersion profiles.
Also, incomplete relaxation or anisotropic velocity distributions might contribute to the failing of the King model.
In the case of BH subsystems, even an anisotropic generalization of the King-Michie model \citep{Michie1963,King1966} may not provide a fully consistent result. 
Currently, the best way to describe an evolved cluster is by a Fokker-Planck (FP) evolutionary model (direct solution, Monte-Carlo or gaseous model).
However, we still can use the underlying physical meaning of the King model (connecting surface density profile and velocity dispersion) to attribute some predictive power to its fitting.

\cite{Rodriguez2015} developed a hybrid Monte-Carlo code which combines the Monte-Carlo method with direct $N$-body calculations of BHs.
They compared the core radii evolution of star clusters with initial small $N$ using their new code, the Monte-Carlo code CMC \citep{Joshi2000} and NBODY6 and found their new approach is consistent with NBODY6, while CMC is not.
In contrast to their results, the comparison of our results with Monte-Carlo (MOCCA) models show a consistent half-mass radius and core radius evolution.
More detailed comparisons between our results and models using CMC will be discussed in future publications (Morscher et al., priv. comm.).

With our million-body models we have not only entered a new phase in the astrophysical modeling of GCs, but also computationally there are some very remarkable observations, which we summarize here (see also \citealp{Wang2015} for details).
The general perception over several decades is that direct $N$-body simulations have a bottleneck in the incredible number of pairwise gravitational force computations (scaling asymptotically as $N^2$). 
Due to the most powerful GPU accelerators (currently we use NVIDIA Kepler architecture) only a few GPUs ($8$ to $16$) are sufficient to make the million-body simulation of GCs feasible.
This work is just the beginning of direct $N$-body simulations of realistic GCs.
The computing times of these four models are $2000$-$4000$ hours per initial half-mass relaxation time $T_{\rm rh0}$ ($0.5$-$2.0~T_{\rm rh0}$ totally).
This estimate has large uncertainty due to the evolution of the relaxation time and the binaries (which cause small time steps).
For a more dense GC with initial half-mass radius $1$~pc, $T_{\rm rh0}$ is about $400$~Myr.
This means in the worst case, $10^5$ hours may be required to finish the $12$~Gyr simulation, if $T_{\rm rh}$ and $N$ do not change during the $12$~Gyr evolution.
Fortunately, $T_{\rm rh}$ increases due to the expansion of GCs and $N$ also decreases because of tidal evaporation.
Hence, the computing time can be reduced significantly at the late stage of GC evolution \citep[see, e.g.,][]{Heggie2014}.
With further improvement of computational and communication hardware there is much room for using more GPUs, increasing the parallelism and approach the billion particle limit in the Exaflop/s scale for direct $N$-body \citep{Huang2015}. 

In future GC simulations, the improvement of stellar evolution recipes is also important.
Currently, NBODY6++GPU cannot handle multiple stellar populations (different metallicities and ages of individual stars) which appear in many GCs (e.g. \citealp{Piotto2009,Gratton2012}). 
NGC~4372 also shows this feature (AGB spread in Fig.~\ref{fig:cmdobs}) and there is a Na-O anti-correlation \citep{San2015}.
The biggest challenge of this task results from the complexity of binary stellar evolution involving two components with different metallicities and ages.
Besides, the stellar evolution is also not fully consistent with observations as there are different CHB features when comparing our models with NGC~4372 (Fig.~\ref{fig:cmdobs}), although these differences could also have a dynamical origin.

The kick models for NSs and BHs significantly influence the dynamical evolution of the GC simulations.
\cite{Podsiadlowski2004} suggest that electron capture supernovae (ECS) can result in reduced kicks of NSs.
\cite{Ivanova2008,Ivanova2014} study the binaries with ECS NSs in GCs and find that the ratio of core collapse supernova NSs to ECS NSs could be $30-200$ to $1$ in a typical GC, as compared to $1$ to $10$ in the field.
In the Monte-Carlo models (e.g. \citealp{Morscher2015}), this kick mechanism is also implemented.
Future (DRAGON) cluster modeling should include this mechanism for NS formation.
This might solve the issue of inconsistency for high kick velocities observed in the Galactic field \citep{Hobbs2005} and the discovery of NS X-ray binaries in GCs \citep{Manchester2005}.

\section*{Acknowledgments}
We acknowledge support through the Silk Road Project at the National Astronomical Observatories of China (NAOC, http://silkroad.bao.ac.cn).

We also acknowledge support by the Max-Planck Computing and Data Facility (MPCDF, http://www.mpcdf.mpg.de/) in Garching, Germany. 
All simulations were run on their ¡°Hydra¡± GPU cluster. 

We are grateful to Nikolay Kacharov and Paolo Bianchini for their help in providing the observation data of NGC~4372 and thank Douglas Heggie for useful suggestions.
We thank Tjibaria Pijloo for the help in the discussion of initial conditions of our models.
We thank Wojciech Pych for helping us develop COCOA and the sim2obs tool which assisted in producing the images in this paper. 

Part of this work has been funded by the "Alexander von Humboldt Polish Honorary Research Fellowship" awarded to R.S. by the Foundation for Polish Science. L.W. and R.S. thank M.G. and the Nicolaus Copernicus Astronomical Center,  Warsaw, Poland for hospitality and support.
R.S. and P.B are grateful for support by the Chinese Academy of Sciences Visiting Professorship for Senior International Scientists, Grant Number 2009S1$-$5, and through the ``Qianren'' special foreign experts program of China, both at the NAOC. 
R.S. and P.B. and L.W. are grateful for kind hospitality and support during several visits to the Max-Planck-Institute for Astrophysics.
S.A. is grateful for support during several visits to the Kavli Institute for Astronomy and Astrophysics, Peking University and NAOC.
M.G. and A.A. acknowledge a partial support from the Polish Ministry of Sciences and Higher Education through the grant DEC-2012/07/B/ST9/04412.
A.A. also acknowledges partial support from Nicolaus Copernicus Astronomical Center's grant for young researchers. 
T.N. acknowledges support by the MPCDF cluster of excellence ¡®Origin and Structure of the Universe¡¯.
M.B.N.K. was supported by  the  Peter  and  Patricia  Gruber  Foundation through the  PPGF  Fellowship,  by  the  Peking  University  One Hundred Talent Fund (985), and by the National Natural Science Foundation  of  China  (grants  11050110414, 11173004, 11573004). 

Finally, we thank the referee, Jarrod Hurley, for many helpful suggestions which helped to improve the manuscript.


\begin{thebibliography}{999}

\bibitem[\protect\citeauthoryear{Aarseth}{2003}]{Aarseth2003}
Aarseth S.~J., 2003, Gravitational N-Body Simulations, Cambridge University Press

\bibitem[\protect\citeauthoryear{Aarseth}{2012}]{Aarseth2012}
Aarseth S.~J., 2012, MNRAS, 422, 841 

\bibitem[\protect\citeauthoryear{Alexander et 
al.}{2014}]{Alexander2014} Alexander P.~E.~R., Gieles M., Lamers 
H.~J.~G.~L.~M., Baumgardt H., 2014, MNRAS, 442, 1265 

\bibitem[\protect\citeauthoryear{Alcaino et al.}{1991}]{Alcaino1991} Alcaino G., Liller W., Alvarado F., Wenderoth E., 1991, AJ, 102, 159 

\bibitem[\protect\citeauthoryear{Askar et al.}{2014}]{Askar2014} Askar A., Giersz M., Pych W., Olech A., Hypki A., 2014, arXiv, arXiv:1501.00417

\bibitem[\protect\citeauthoryear{Baumgardt \& Makino}{2003}]{Baumgardt2003} Baumgardt H., Makino J., 2003, MNRAS, 340, 227 

\bibitem[\protect\citeauthoryear{Banerjee, Baumgardt, 
\& Kroupa}{2010}]{Banerjee2010} Banerjee S., Baumgardt H., Kroupa P., 2010, MNRAS, 402, 371 

\bibitem[\protect\citeauthoryear{Belczynski, Kalogera, 
\& Bulik}{2002}]{Belczynski2002} Belczynski K., Kalogera V., Bulik T., 2002, ApJ, 572, 407 

\bibitem[\protect\citeauthoryear{Breen 
\& Heggie}{2013}]{Breen2013} Breen P.~G., Heggie D.~C., 2013, MNRAS, 432, 2779 

\bibitem[\protect\citeauthoryear{Brodie 
\& Strader}{2006}]{Brodie2006} Brodie J.~P., Strader J., 2006, ARA\&A, 44, 193 

\bibitem[\protect\citeauthoryear{Casertano 
\& Hut}{1985}]{Casertano1985} Casertano S., Hut P., 1985, ApJ, 298, 80 

\bibitem[\protect\citeauthoryear{Casetti-Dinescu et 
al.}{2007}]{Casetti2007} Casetti-Dinescu D.~I., Girard T.~M., 
Herrera D., van Altena W.~F., L{\'o}pez C.~E., Castillo D.~J., 2007, AJ, 
134, 195 

\bibitem[\protect\citeauthoryear{Chatterjee et
al.}{2013}]{Chatterjee2013} Chatterjee S., Umbreit S., Fregeau J.~M.,
Rasio F.~A., 2013, MNRAS, 429, 2881

\bibitem[\protect\citeauthoryear{Contenta, Varri, 
\& Heggie}{2015}]{Contenta2015} Contenta F., Varri A.~L., Heggie D.~C., 2015, MNRAS, 449, L100 

\bibitem[\protect\citeauthoryear{Davies, Piotto, 
\& de Angeli}{2004}]{Davies2004} Davies M.~B., Piotto G., de Angeli F., 2004, MNRAS, 349, 129 

\bibitem[\protect\citeauthoryear{Davies}{2015}]{Davies2015} Davies 
M.~B., 2015, Ecology of Blue Straggler StarsAstrophysics and Space Science Library, Volume 413.~ISBN 978-3-662-44433-7.~Springer-Verlag Berlin Heidelberg, 2015, p.~203, 203 

\bibitem[\protect\citeauthoryear{Downing}{2012}]{Downing2012} 
Downing J.~M.~B., 2012, MNRAS, 425, 2234 

\bibitem[\protect\citeauthoryear{Eggleton, Fitchett, \& Tout}{1989}]{Eggleton1989} Eggleton P.~P., Fitchett M.~J., Tout C.~A., 1989, ApJ, 347, 998 

\bibitem[\protect\citeauthoryear{Einsel 
\& Spurzem}{1999}]{Einsel1999} Einsel C., Spurzem R., 1999, MNRAS, 302, 81 

\bibitem[\protect\citeauthoryear{Fregeau et 
al.}{2003}]{Fregeau2003} Fregeau J.~M., G{\"u}rkan M.~A., Joshi 
K.~J., Rasio F.~A., 2003, ApJ, 593, 772 

\bibitem[\protect\citeauthoryear{Fregeau et al.}{2004}]{Fregeau2004} 
Fregeau J.~M., Cheung P., Portegies Zwart S.~F., Rasio F.~A., 2004, MNRAS, 352, 1

\bibitem[\protect\citeauthoryear{Fregeau 
\& Rasio}{2007}]{Fregeau2007} Fregeau J.~M., Rasio F.~A., 2007, ApJ, 658, 1047 

\bibitem[\protect\citeauthoryear{Fryer}{2004}]{Fryer2004} Fryer 
C.~L., 2004, ApJ, 601, L175 

\bibitem[\protect\citeauthoryear{Fukushige 
\& Heggie}{2000}]{Fukushige2000} Fukushige T., Heggie D.~C., 2000, MNRAS, 318, 753 

\bibitem[\protect\citeauthoryear{Gaburov, Harfst \& Portegies Zwart}{2009}]{Gaburov2009} Gaburov E., Harfst S., Portegies Zwart S., 2009, NewA, 14, 630 

\bibitem[\protect\citeauthoryear{Geisler et 
al.}{1995}]{Geisler1995} Geisler D., Piatti A.~E., Claria J.~J., 
Minniti D., 1995, AJ, 109, 605 

\bibitem[\protect\citeauthoryear{Genzel, Eisenhauer, 
\& Gillessen}{2010}]{Genzel2010} Genzel R., Eisenhauer F., Gillessen S., 2010, RvMP, 82, 3121 

\bibitem[\protect\citeauthoryear{Giersz 
\& Heggie}{1994}]{Giersz1994} Giersz M., Heggie D.~C., 1994, MNRAS, 268, 257 

\bibitem[\protect\citeauthoryear{Giersz}{2006}]{Giersz2006} Giersz 
M., 2006, MNRAS, 371, 484 

\bibitem[\protect\citeauthoryear{Giersz
\& Heggie}{2011}]{Giersz2011} Giersz M., Heggie D.~C., 2011, MNRAS, 410, 2698

\bibitem[\protect\citeauthoryear{Gieles, Heggie, 
\& Zhao}{2011}]{Gieles2011} Gieles M., Heggie D.~C., Zhao H., 2011, MNRAS, 413, 2509 

\bibitem[\protect\citeauthoryear{Giersz et al.}{2013}]{Giersz2013} 
Giersz M., Heggie D.~C., Hurley J.~R., Hypki A., 2013, MNRAS, 431, 2184 

\bibitem[\protect\citeauthoryear{Giersz et al.}{2015}]{Giersz2015} Giersz M., Leigh N., Hypki A., L{\"u}tzgendorf N., Askar A., 2015, MNRAS, 454, 3150 

\bibitem[\protect\citeauthoryear{Gratton, Carretta, 
\& Bragaglia}{2012}]{Gratton2012} Gratton R.~G., Carretta E., Bragaglia A., 2012, A\&ARv, 20, 50 

\bibitem[\protect\citeauthoryear{Harris}{1996}]{Harris1996} Harris 
W.~E., 1996, AJ, 112, 1487 

\bibitem[\protect\citeauthoryear{Heggie}{1975}]{Heggie1975} Heggie 
D.~C., 1975, MNRAS, 173, 729 

\bibitem[\protect\citeauthoryear{Heggie, Trenti, 
\& Hut}{2006}]{Heggie2006} Heggie D.~C., Trenti M., Hut P., 2006, MNRAS, 368, 677 

\bibitem[\protect\citeauthoryear{Heggie
\& Giersz}{2008}]{Heggie2008} Heggie D.~C., Giersz M., 2008, MNRAS, 389, 1858

\bibitem[\protect\citeauthoryear{Heggie 
\& Giersz}{2009}]{Heggie2009} Heggie D.~C., Giersz M., 2009, MNRAS, 397, L46 

\bibitem[\protect\citeauthoryear{Heggie}{2014}]{Heggie2014} Heggie 
D.~C., 2014, MNRAS, 445, 3435 

\bibitem[\protect\citeauthoryear{H{\'e}non}{1971}]{Henon1971} H{\'e}non M., 1971, Ap\&SS, 13, 284 

\bibitem[\protect\citeauthoryear{Hills}{1975}]{Hills1975} Hills 
J.~G., 1975, AJ, 80, 809 

\bibitem[\protect\citeauthoryear{Hobbs et al.}{2005}]{Hobbs2005}
Hobbs G., Lorimer D.~R., Lyne A.~G., Kramer M., 2005, MNRAS, 360, 974 

\bibitem[\protect\citeauthoryear{Huang, Spurzem, 
\& Berczik}{2015}]{Huang2015} Huang S., Spurzem R., Berczik P., 2015, arXiv, arXiv:1508.02510 

\bibitem[\protect\citeauthoryear{Hurley, Pols, 
\& Tout}{2000}]{Hurley2000} Hurley J.~R., Pols O.~R., Tout C.~A., 2000, MNRAS, 315, 543 

\bibitem[\protect\citeauthoryear{Hurley, Tout, 
\& Pols}{2002}]{Hurley2002} Hurley J.~R., Tout C.~A., Pols O.~R., 2002, MNRAS, 329, 897 

\bibitem[\protect\citeauthoryear{Hurley et al.}{2005}]{Hurley2005} 
Hurley J.~R., Pols O.~R., Aarseth S.~J., Tout C.~A., 2005, MNRAS, 363, 293 

\bibitem[\protect\citeauthoryear{Hurley}{2007}]{Hurley2007} Hurley 
J.~R., 2007, MNRAS, 379, 93 

\bibitem[\protect\citeauthoryear{Hurley 
\& Shara}{2012}]{Hurley2012} Hurley J.~R., Shara M.~M., 2012, MNRAS, 425, 2872 

\bibitem[\protect\citeauthoryear{Hypki 
\& Giersz}{2013}]{Hypki2013} Hypki A., Giersz M., 2013, MNRAS, 429, 1221 

\bibitem[\protect\citeauthoryear{Hong et al.}{2013}]{Hong2013} 
Hong J., Kim E., Lee H.~M., Spurzem R., 2013, MNRAS, 430, 2960 

\bibitem[\protect\citeauthoryear{Iwasawa, Portegies Zwart, 
\& Makino}{2015}]{Iwasawa2015} Iwasawa M., Portegies Zwart S., Makino J., 2015, ComAC, 2, 6 

\bibitem[\protect\citeauthoryear{Ivanova et al.}{2008}]{Ivanova2008} Ivanova N., Heinke C.~O., Rasio F.~A., Belczynski K., Fregeau J.~M., 2008, MNRAS, 386, 553 

\bibitem[\protect\citeauthoryear{Ivanova}{2014}]{Ivanova2014} Ivanova, N., in ``Star clusters and black holes in galaxies across cosmic time Proceedings IAU Symposium No. 312'', 2014; Y. Meiron, S. Li, F.-K. Liu \& R. Spurzem (eds.), doi:10.1017/S1743921315007826

\bibitem[\protect\citeauthoryear{Lane et al.}{2009}]{Lane2009} 
Lane R.~R., Kiss L.~L., Lewis G.~F., Ibata R.~A., Siebert A., Bedding 
T.~R., Sz{\'e}kely P., 2009, MNRAS, 400, 917 

\bibitem[\protect\citeauthoryear{Joshi, Rasio, 
\& Portegies Zwart}{2000}]{Joshi2000} Joshi K.~J., Rasio F.~A., Portegies Zwart S., 2000, ApJ, 540, 969 

\bibitem[\protect\citeauthoryear{Kacharov et 
al.}{2014}]{Kacharov2014} Kacharov N., et al., 2014, A\&A, 567, A69 

\bibitem[\protect\citeauthoryear{Kim et al.}{2008}]{Kim2008} 
Kim E., Yoon I., Lee H.~M., Spurzem R., 2008, MNRAS, 383, 2 

\bibitem[\protect\citeauthoryear{King}{1962}]{King1962} King I., 
1962, AJ, 67, 471 

\bibitem[\protect\citeauthoryear{Kouwenhoven et 
al.}{2007}]{Kouwenhoven2007} Kouwenhoven M.~B.~N., Brown A.~G.~A., Portegies Zwart S.~F., Kaper L., 2007, A\&A, 474, 77 

\bibitem[\protect\citeauthoryear{Kroupa, Tout 
\& Gilmore}{1993}]{Kroupa1993} Kroupa P., Tout C.~A., Gilmore G., 1993, MNRAS, 262, 545 

\bibitem[\protect\citeauthoryear{Kroupa}{1995}]{Kroupa1995} Kroupa 
P., 1995, MNRAS, 277, 1491 

\bibitem[\protect\citeauthoryear{Kroupa}{2001}]{Kroupa2001} Kroupa 
P., 2001, MNRAS, 322, 231 

\bibitem[\protect\citeauthoryear{Kruijssen et 
al.}{2011}]{Kruijssen2011} Kruijssen J.~M.~D., Pelupessy F.~I., 
Lamers H.~J.~G.~L.~M., Portegies Zwart S.~F., Icke V., 2011, MNRAS, 414, 
1339 

\bibitem[\protect\citeauthoryear{Kustaanheimo \& Stiefel}{1965}]{Kustaanheimo1965}
Kustaanheimo P., Stiefel E., 1965, J.~Reine Angew.~Math., 218, 204

\bibitem[\protect\citeauthoryear{K{\"u}pper et 
al.}{2010}]{Kupper2010} K{\"u}pper A.~H.~W., Kroupa P., Baumgardt 
H., Heggie D.~C., 2010, MNRAS, 407, 2241 

\bibitem[\protect\citeauthoryear{King}{1966}]{King1966} King 
I.~R., 1966, AJ, 71, 64 

\bibitem[\protect\citeauthoryear{Lamers, Gieles, 
\& Portegies Zwart}{2005}]{Lamers2005} Lamers H.~J.~G.~L.~M., Gieles M., Portegies Zwart S.~F., 2005, A\&A, 429, 173 

\bibitem[\protect\citeauthoryear{Leigh et al.}{2013}]{Leigh2013} 
Leigh N., Giersz M., Webb J.~J., Hypki A., De Marchi G., Kroupa P., Sills 
A., 2013, MNRAS, 436, 3399 

\bibitem[\protect\citeauthoryear{Leigh et al.}{2015}]{Leigh2015} 
Leigh N.~W.~C., Giersz M., Marks M., Webb J.~J., Hypki A., Heinke C.~O., 
Kroupa P., Sills A., 2015, MNRAS, 446, 226 

\bibitem[\protect\citeauthoryear{Mackey et al.}{2008}]{Mackey2008} 
Mackey A.~D., Wilkinson M.~I., Davies M.~B., Gilmore G.~F., 2008, MNRAS, 
386, 65 

\bibitem[\protect\citeauthoryear{Makino 
\& Hut}{1988}]{Makino1988} Makino J., Hut P., 1988, ApJS, 68, 833 

\bibitem[\protect\citeauthoryear{Makino}{1996}]{Makino1996} Makino 
J., 1996, ApJ, 471, 796 

\bibitem[\protect\citeauthoryear{Makino et al.}{2003}]{Makino2003}
Makino J., Fukushige T., Koga M., Namura K., 2003, PASJ, 55, 1163 

\bibitem[\protect\citeauthoryear{Manchester et 
al.}{2005}]{Manchester2005} Manchester R.~N., Hobbs G.~B., Teoh A., 
Hobbs M., 2005, AJ, 129, 1993 

\bibitem[\protect\citeauthoryear{Merritt et 
al.}{2004}]{Merritt2004} Merritt D., Piatek S., Portegies Zwart S., 
Hemsendorf M., 2004, ApJ, 608, L25 

\bibitem[\protect\citeauthoryear{Michie}{1963}]{Michie1963} Michie 
R.~W., 1963, MNRAS, 125, 127 

\bibitem[\protect\citeauthoryear{Mikkola 
\& Aarseth}{1993}]{Mikkola1993} Mikkola S., Aarseth S.~J., 1993, CeMDA, 57, 439 

\bibitem[\protect\citeauthoryear{Moffat}{1969}]{Moffat1969} Moffat A.~F.~J., 1969, A\&A, 3, 455 

\bibitem[\protect\citeauthoryear{Morscher et 
al.}{2015}]{Morscher2015} Morscher M., Pattabiraman B., Rodriguez C., Rasio F.~A., Umbreit S., 2015, ApJ, 800, 9 

\bibitem[\protect\citeauthoryear{Nitadori 
\& Aarseth}{2012}]{Nitadori2012} Nitadori K., Aarseth S.~J., 2012, MNRAS, 424, 545 

\bibitem[\protect\citeauthoryear{Noyola 
\& Gebhardt}{2006}]{Noyola2006} Noyola E., Gebhardt K., 2006, AJ, 132, 447 

\bibitem[\protect\citeauthoryear{Peng et al.}{2011}]{Peng2011} 
Peng E.~W., et al., 2011, ApJ, 730, 23 

\bibitem[\protect\citeauthoryear{Podsiadlowski et al.}{2004}]{Podsiadlowski2004} Podsiadlowski P., Langer N., Poelarends A.~J.~T., Rappaport S., Heger A., Pfahl E., 2004, ApJ, 612, 1044 

\bibitem[\protect\citeauthoryear{Pijloo et al.}{2015}]{Pijloo2015} Pijloo J.~T., Portegies Zwart S.~F., Alexander P.~E.~R., Gieles M., Larsen S.~S., Groot P.~J., Devecchi B., 2015, MNRAS, 453, 605 

\bibitem[\protect\citeauthoryear{Piotto et 
al.}{2002}]{Piotto2002} Piotto G., et al., 2002, A\&A, 391, 945 

\bibitem[\protect\citeauthoryear{Piotto}{2009}]{Piotto2009} Piotto 
G., 2009, IAUS, 258, 233 

\bibitem[\protect\citeauthoryear{Portegies Zwart, Hut, 
\& Verbunt}{1997}]{PZ1997} Portegies Zwart S.~F., Hut P., Verbunt F., 1997, A\&A, 328, 130 

\bibitem[\protect\citeauthoryear{Portegies Zwart et 
al.}{2004}]{PZ2004} Portegies Zwart S.~F., Baumgardt H., Hut 
P., Makino J., McMillan S.~L.~W., 2004, Natur, 428, 724 

\bibitem[\protect\citeauthoryear{Rodriguez et al.}{2015}]{Rodriguez2015} Rodriguez C.~L., Pattabiraman B., 
Chatterjee S., Choudhary A., Liao W.-k., Morscher M., Rasio F.~A., 2015, 
arXiv, arXiv:1511.00695 

\bibitem[\protect\citeauthoryear{San Roman et 
al.}{2015}]{San2015} San Roman I., et al., 2015, A\&A, 579, A6 

\bibitem[\protect\citeauthoryear{Sippel 
\& Hurley}{2013}]{Sippel2013} Sippel A.~C., Hurley J.~R., 2013, MNRAS, 430, L30 

\bibitem[\protect\citeauthoryear{Spitzer}{1987}]{Spitzer1987} 
Spitzer L., 1987, Dynamical evolution of globular clusters, Princeton University Press

\bibitem[\protect\citeauthoryear{Spurzem 
\& Aarseth}{1996}]{Spurzem1996} Spurzem R., Aarseth S.~J., 1996, MNRAS, 282, 19 

\bibitem[\protect\citeauthoryear{Stodolkiewicz}{1986}]{Stodolkiewicz1986} 
Stodolkiewicz J.~S., 1986, AcA, 36, 19 

\bibitem[\protect\citeauthoryear{Sugimoto et 
al.}{1990}]{Sugimoto1990} Sugimoto D., Chikada Y., Makino J., Ito 
T., Ebisuzaki T., Umemura M., 1990, Natur, 345, 33 

\bibitem[\protect\citeauthoryear{Trager, King, 
\& Djorgovski}{1995}]{Trager1995} Trager S.~C., King I.~R., Djorgovski S., 1995, AJ, 109, 218 

\bibitem[\protect\citeauthoryear{Wang et al.}{2015}]{Wang2015} 
Wang L., Spurzem R., Aarseth S., Nitadori K., Berczik P., Kouwenhoven 
M.~B.~N., Naab T., 2015, MNRAS, 450, 4070 

\bibitem[\protect\citeauthoryear{Zonoozi et 
al.}{2011}]{Zonoozi2011} Zonoozi A.~H., K{\"u}pper A.~H.~W., 
Baumgardt H., Haghi H., Kroupa P., Hilker M., 2011, MNRAS, 411, 1989 

\bibitem[\protect\citeauthoryear{Zonoozi et 
al.}{2014}]{Zonoozi2014} Zonoozi A.~H., Haghi H., K{\"u}pper 
A.~H.~W., Baumgardt H., Frank M.~J., Kroupa P., 2014, MNRAS, 440, 3172 

\end{thebibliography}
\end{document}